\documentclass[twocolumn,secnumarabic,superscriptaddress,amssymb, nobibnotes, aps, prb]{revtex4-2}

\usepackage{amsmath}
\usepackage{mathrsfs}  % use to write fancy letters (e.g. for free energy)
\usepackage{mathtools} % use to write := im math
\usepackage{bbold}    % use to write fancy 1 for unity matrix
\usepackage{braket}    % lets you use the bra, ket notation
\usepackage{gensymb}   % for \degree

\usepackage{graphicx}      	% import figures
\usepackage{multirow}       % merge rows and columns in table
\usepackage{here} 			% display figures exactly where they are in code
\usepackage{longtable}

\usepackage{url}		% display urls in a nice way
\usepackage[plainpages=false,pdfpagelabels=true, linkcolor=cyan,breaklinks,colorlinks=true]{hyperref}  % hyperrefs
\usepackage{soul}                   % allows to cross out text
\usepackage[dvipsnames]{xcolor}		% text in different colors

\usepackage{siunitx}
\usepackage{xspace}

\newcommand{\ute}{UTe\textsubscript{2}\xspace}
\newcommand{\mnge}{Mn\textsubscript{3}Ge\xspace}
\newcommand{\sto}{SrTiO\textsubscript{3}\xspace}
\newcommand{\sro}{Sr\textsubscript{2}RuO\textsubscript{4}\xspace}

\begin{document}

\title{Resonant Ultrasound Spectroscopy for Irregularly-Shaped Samples and its Application to Uranium Ditelluride}%

\author{Florian Theuss}%
\affiliation{Laboratory of Atomic and Solid State Physics, Cornell University, Ithaca, NY 14853, USA}
\author{Gregorio de la Fuente Simarro}
\affiliation{Laboratory of Atomic and Solid State Physics, Cornell University, Ithaca, NY 14853, USA}
\author{Avi Shragai}
\affiliation{Laboratory of Atomic and Solid State Physics, Cornell University, Ithaca, NY 14853, USA}
\author{Gael Grissonnanche}
\affiliation{Laboratory of Atomic and Solid State Physics, Cornell University, Ithaca, NY 14853, USA}
\author{Ian M Hayes}
\affiliation{Quantum Materials Center, Department of Physics,University of Maryland, College Park, Maryland 20742, USA}
\author{Shanta Saha}
\affiliation{Quantum Materials Center, Department of Physics,University of Maryland, College Park, Maryland 20742, USA}
\author{Tatsuya Shishidou}
\affiliation{Department of Physics, University of Wisconsin-Milwaukee, Milwaukee, Wisconsin 53201, USA}
\author{Taishi Chen}
\affiliation{The Institute for Solid State Physics, The University of Tokyo, Kashiwa, Chiba 277-8581, Japan}
\author{Satoru Nakatsuji}
\affiliation{The Institute for Solid State Physics, The University of Tokyo, Kashiwa, Chiba 277-8581, Japan}
\affiliation{Department of Physics, The University of Tokyo, Tokyo  113-0033, Japan}
\affiliation{Institute for Quantum Matter and Department of Physics and Astronomy, Johns Hopkins University, Baltimore, MD 21218, USA}
\affiliation{Trans-scale Quantum Science Institute, University of Tokyo, Tokyo 113-0033, Japan}
\author{Sheng Ran}
\affiliation{Department of Physics, Washington University in St. Louis, St. Louis, MO 63130, USA}
\author{Michael Weinert}
\affiliation{Department of Physics, University of Wisconsin-Milwaukee, Milwaukee, Wisconsin 53201, USA}
\author{Nicholas P Butch}
\affiliation{Quantum Materials Center, Department of Physics,University of Maryland, College Park, Maryland 20742, USA}
\affiliation{NIST Center for Neutron Research, National Institute of Standards and Technology, 100 Bureau Drive, Gaithersburg, Maryland 20899, USA}
\author{Johnpierre Paglione}
\affiliation{Quantum Materials Center, Department of Physics,University of Maryland, College Park, Maryland 20742, USA}
\author{B. J. Ramshaw}
\email{bradramshaw@cornell.edu}
\affiliation{Laboratory of Atomic and Solid State Physics, Cornell University, Ithaca, NY 14853, USA}

\date{\today}%

\begin{abstract}
Resonant ultrasound spectroscopy (RUS) is a powerful technique for measuring the full elastic tensor of a given material in a single experiment. Previously, this technique was limited to regularly-shaped samples such as rectangular parallelepipeds, spheres, and cylinders \cite{visscher_normal_1991}. We demonstrate a new method for determining the elastic moduli of irregularly-shaped samples, extending the applicability of RUS to a much larger set of materials. We apply this new approach to the recently-discovered unconventional superconductor \ute and provide its elastic tensor at both 300 and 4 kelvin.
\end{abstract}

\maketitle
% \tableofcontents

%%%%%%%%%%%%% input introduction
\section{Introduction}
\ute is a recently-discovered unconventional superconductor \cite{ranNearlyFerromagneticSpintriplet2019} with large upper critical fields \cite{aokiUnconventionalSuperconductivityHeavy2019, ranNearlyFerromagneticSpintriplet2019} and an NMR Knight shift \cite{nakamineSuperconductingPropertiesHeavy2019, nakamineInhomogeneousSuperconductingState2021, fujibayashiSuperconductingOrderParameter2022} that suggest spin-triplet pairing. There are, however, many unsolved mysteries in \ute, including field-reentrant superconductivity \cite{knebelFieldreentrantSuperconductivityClose2019, ranExtremeMagneticFieldboosted2019, ranEnhancementReentranceSpin2020, aokiMultipleSuperconductingPhases2020}, time-reversal symmetry breaking \cite{weiInterplayMagnetismSuperconductivity2021,hayesMulticomponentSuperconductingOrder2021}, a phase transition between superconducting ground states as a function of magnetic field \cite{rosuel2022FieldinducedTuningPairing}, and the occurrence of two superconducting transitions in certain samples \cite{Cairns2020, hayesMulticomponentSuperconductingOrder2021, thomasSpatiallyInhomogeneousSuperconductivity2021, rosaSinglecomponentSuperconductingState2022}. Underlying many of these mysteries is the question of what is the symmetry of the superconducting order parameter in \ute.

Externally applied stress has proven to be a useful tuning parameter when investigating these types of questions. For example, there are two unambiguous superconducting phase transitions in \ute under hydrostatic pressure \cite{braithwaiteMultipleSuperconductingPhases2019, aokiMultipleSuperconductingPhases2020}, and uniaxial pressure experiments \cite{girod2022ThermodynamicElectricalTransport} imply an insensitivity of the superconducting order parameter to shear strain. However, while stress and pressure are conceptually straightforward parameters to tune externally, the more physically-relevant quantity---related to microscopic bond distances and unit cell volumes---is strain. Stress, $\sigma$, and strain, $\varepsilon$, are linearly related through the elastic tensor, $\sigma = \boldsymbol{c} \varepsilon$, and converting from the experimentally-determined stress to strain requires the full elastic tensor. For example, the full elastic tensor was central in determining the quantitative relationship between strain, the van Hove point, and superconductivity in \sro \cite{barber2019role,li2022elastocaloric}, as well as the relationship between anisotropic strains and superconductivity in CeIrIn$_5$ \cite{bachmann2019spatial}.

The full elastic tensor of a material can be determined with resonant ultrasound spectroscopy (RUS). RUS measures the mechanical resonance spectrum of a three dimensional solid. The resonance frequencies are determined by both intrinsic sample properties---the density and elastic moduli---as well as by the sample boundary conditions. If the elastic moduli, density, and sample geometry are known, then the resonance frequencies are easily calculated numerically---the ``forward problem''---either using the method of \citet{visscher_normal_1991} or by finite elements \cite{liu2012measuring,plesek2004using}. The inverse problem---obtaining the elastic moduli from a measured resonance spectrum---is more challenging because it requires multiple numerical evaluations of the forward problem and the navigation of a complex parameter landscape with many local minima \cite{Ramshaw2015a}. 

For simple geometries with easily-defined boundaries---typically rectangular parallelepipeds, cylinders, or spheres---the method of \citet{visscher_normal_1991} can be combined with either Levenberg-Marquardt or heuristic (such as genetic algorithm) fitting methods to solve the inverse problem and obtain the elastic moduli from the resonance spectrum. For irregular samples, however, finite elements has been traditionally used to solve the forward problem. The difficulty with this approach is that finite elements is too computationally intensive to use when solving the inverse problem. This has largely restricted the applicability of RUS to materials where regularly-shaped samples can be prepared.

In the case of \ute, preparation of a rectangular parallelepiped is difficult due to the brittle nature of the material (in addition to the potential hazards associated with polishing uranium compounds). In this work, we take advantage of the recent development by \citet{shragai2023RapidMethodComputing} that solves the forward problem for irregularly-shaped samples in a way that is two orders of magnitude faster than finite elements. This has allowed us to develop a protocol for solving the inverse problem for irregularly-shaped samples. We demonstrate this protocol on samples of \sto and \mnge---compounds with known elastic moduli---and then apply our new technique to obtain the full elastic tensor of \ute at 300~K and at 4 K. 

\section{Methods}

\begin{figure}%[H]
	\includegraphics[width=1\columnwidth]{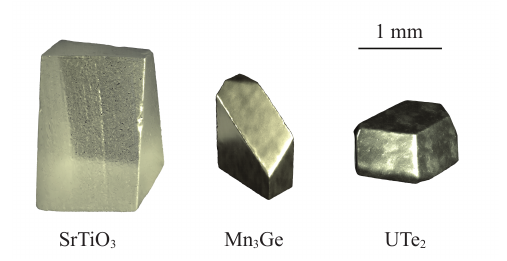}
		\caption{\textbf{CT-Scan Models.} 3D models of the irregularly-shaped samples used for RUS measurements based on CT-scans. From left to right: \sto sample B, \mnge sample B, \ute.}
	\label{fig:sketch}
\end{figure}

\textbf{Resonant ultrasound spectroscopy.} We performed RUS measurements by placing a sample in weak mechanical contact between two piezoelectric transducers, providing nearly-free boundary conditions. One transducer is driven with an AC voltage at fixed frequency, and the voltage generated on the other transducer is measured using a lockin amplifier. By stepping the drive frequency from roughly 100 kHz to 5 MHz, we obtain the first one hundred or so mechanical resonances for a typical, mm-scale sample. More details of the experimental setup and how to measure resonance spectra can be found in \citet{Ramshaw2015a} and \citet{balakirev2019resonant}. Full lists of all experimental resonance spectra used in this study are given in the supplement. 

%\flo{Commonly, the resonance spectra are calculated} according to earlier work by Visscher et al. \cite{visscher_normal_1991}, minimizing the linear elastic Lagrangian as a function of the displacement field $\vec{u} \left( \vec{r}, t \right)$
%\begin{equation}
	%\label{eq:Lagrangian}
	%\mathcal{L} = \frac{1}{2} \int \left[\sum_i \rho \dot{u}_i (\vec{r}) - \frac{1}{2} \sum_{ijkl} c_{ijkl} \frac{\partial u_i(\vec{r})}{r_j} \frac{\partial u_k(\vec{r})}{r_l}\right] dV,
%\end{equation}
%where $\rho$ is the density, $c_{ijkl}$ is the elastic tensor, and the integral is over the entire volume of the measured sample. Due to the complexity of the volume integral, this analysis has almost exclusively been done by expanding the displacement field in Cartesian polynomials and preparing the measured samples in simple geometries such as rectangular parallelepipeds, cylinders, or spheres \cite{Ramshaw2015a, Ghosh2020,balakirev2019resonant,Ghosh2020b,Theuss2022}.

\textbf{Fitting resonance spectra.} Resonance spectra are calculated by minimizing the linear elastic Lagrangian as a function of the displacement field $\vec{u} \left( \vec{r}, t \right)$,
\begin{equation}
	\label{eq:Lagrangian}
	\mathcal{L} = \frac{1}{2} \int \left(\sum_i \rho \dot{u}_i^2 (\vec{r}) - \sum_{ijkl} c_{ijkl} \frac{\partial u_i(\vec{r})}{\partial r_j} \frac{\partial u_k(\vec{r})}{\partial r_l}\right) dV,
\end{equation}
where $\rho$ is the density, $c_{ijkl}$ is the elastic tensor, and the integral is over the entire volume of the sample. The widely-adopted method of \citet{visscher_normal_1991} expands the displacements fields in a polynomial basis and solves the volume integrals numerically. This method works well for rectangular parallelepipeds, cylinders, and spheres \cite{balakirev2019resonant}, where the boundaries are easily defined. For irregularly-shaped samples, however, this method is insufficient and new methods for computing the resonance spectra must be used.

We implement two new resonance spectra calculation methods---two ``forward'' solvers---into a genetic algorithm, extending the RUS fitting routine to irregularly-shaped samples. Details of the genetic algorithm itself can be found in \citet{Ramshaw2015a} and in the supplement; here we focus on incorporating and verifying the new forward solvers. 

The first forward solver extends Visscher's method to irregular samples by converting the volume integrals to surface integrals \cite{shragai2023RapidMethodComputing}. We will refer to fits performed with this forward solver as SMI (surface mesh integration \cite{shragai2023RapidMethodComputing}). 

The second forward solver is a commercially-available finite element solver: Comsol. We implement a Comsol forward sovler to verify the newer SMI method (note that the use of FEM to calculate the forward problem for RUS was described earlier in \citet{liu2012measuring}). We will refer to fits performed using Comsol as FEM. 

Both SMI and FEM-based fits are compared to fits using Visscher's approach for rectangular parallelepiped samples. Fits with Visscher's approach are referred to as RPR.

\textbf{Sample digitization and alignment.} Both the SMI and FEM methods require three-dimensional digitizations of the samples. FEM uses the full, three-dimensional tetrahedral mesh of the entire sample volume. SMI uses only the surface triangles of the same mesh (this includes any ``interior'' surfaces around voids). These digitizations were obtained with a Zeiss Xradia Versa XRM-520 X-ray nano-CT and are shown in \autoref{fig:sketch}. The mesh size used for our fits depends on the solver method, as well as on the sample shape and size. We use a mesh with average linear dimension 10~$\mu$m for SMI, and  60~$\mu$m for FEM (FEM uses a larger mesh because it is much more computationally intensive than SMI). The samples are oriented to within 1 degree using Laue backreflection diffractometry. The mesh is then aligned to the crystal axes by identifying two flat faces on the sample which uniquely relate the surface mesh to the orientation of the sample in the Laue apparatus. More details on sample digitization can be found in the SI.

\textbf{Measurement uncertainty.} The uncertainties in our fits have two dominant contributions. One contribution originates from small deviations from the weak-coupling approximation: the resonance frequencies depending weakly on how the sample is mounted in the apparatus. This is predominantly due to the finite force exerted on the sample by the weight of the cantilever holding the top ultrasonic transducer. By remounting the samples in different orientations in our RUS apparatus, we find uncertainties in the elastic moduli of approximately 2~GPa for the \sto and \mnge samples, and 0.5~GPa for \ute.
% All of our \sto and \mnge measurements were performed with levers weighing on the order of 5~g, leading to uncertainties in the elastic moduli of \flo{approximately 2~GPa.} The \ute resonances were measured in a different apparatus with a lighter cantilever (weighing less than 0.5~g), resulting in uncertainties in the elastic moduli of approximately \flo{0.5~GPa}. 
The second largest source of uncertainty is due to uncertainty in the alignment between the crystal axes and the sample mesh. This uncertainty is approximately 1 degree around each axis and results in uncertainties of typically 0.03~GPa for all regularly-shaped samples. This uncertainty is larger for all irregularly-shaped samples---on average 0.4~GPa, 2~GPa, 0.5~GPa for the irregular \sto, \mnge, and \ute samples, respectively. A detailed error analysis can be found in the SI.

\textbf{Density functional theory calculations.} We used density-functional theory to produce estimates of the elastic moduli of \ute. This involved total energy calculations following the procedure of Ravindran et al. \cite{doi:10.1063/1.368733}. The full-potential linearized augmented plane wave method \cite{FLAPW2009} calculations employed the generalized gradient approximation \cite{PBE} for the exchange correlation, wave function and potential energy cutoffs of 16 and 200 Ry, respectively, muffin-tin sphere radii of 1.35 \AA{}, and an $8\times 8\times 8$ $k$-point mesh. Spin-orbit coupling was fully taken into account in the assumed nonmagnetic state. The elastic tensor was extracted from fits of the total energy variations around the experimental structure \cite{UTe2-lat:2006} to the energy-strain formula \cite{doi:10.1063/1.368733}, including linear terms. The resulting moduli are given in \autoref{table:ute2 fit results}.

\section{Test Results}

We test the implementation of two forward solvers---FEM and SMI---for fitting the RUS spectra of irregularly-shaped samples using a genetic algorithm. We compare these results to the moduli extracted for rectangular parallelepiped samples that can be fit using the RPR forward solver, in addition to the FEM and SMI methods. We find agreement between all methods and all sample geometries to within our measurement uncertainties. It's worth noting that while both FEM and SMI are capable of fitting RUS spectra from irregularly-shaped samples, the SMI method is two orders of magnitude faster than FEM, taking under an hour to produce a fit while FEM takes several days or even weeks.

\textbf{\sto.} Our first test system for our new fitting method is \sto, whose elastic tensor consists of only three independent elements due to its cubic crystal structure (point group $O_h$). RUS measurements and fits are performed on two samples (\autoref{table:sto fit results}): Sample A is was polished into a rectangular parallelepiped with dimensions $\left(1.49 \times 2.035 \times 3.02 \right)$~mm$^3$, with edges oriented along the crystallographic axes. We can perform fits using all three methods---RPR, SMI, and FEM---in this simple geometry. Sample B was prepared with an irregular shape (see \autoref{fig:sketch} for a 3D model based on a CT scan). Only SMI and FEM fits are possible for this geometry. 

The fit results are shown in \autoref{table:sto fit results}. Comparing the fits within each sample, we see that all methods yield identical results within uncertainties ($\pm 2$ GPa). This demonstrates that, given the same experimental data and sample geometry, all three forward solvers used in a genetic algorithm yield the same elastic moduli. This is consistent with previous demonstrations that the the three methods of forward computation---RPR, SMI, and FEM---are consistent to better than 1 part in 10$^4$ \cite{shragai2023RapidMethodComputing}.

We observe differences of less than 2.5~\% for all elastic moduli when comparing the rectangular parallelepiped (sample A) and irregular (sample B) samples. These difference are on the order of our estimated uncertainties due to sample loading; alternatively, these deviations could be attributed to systematic uncertainties in the RUS measurement process such as deviations from a perfect rectangular parallelepiped for sample A, or sample imperfections that were not captured by the CT scan such as small voids or inclusions. Within our uncertainty, all of our fits are compatible with the measurements of \citet{bell_elastic_1963} and \citet{poindexter_elastic_1958} (bottom rows of \autoref{table:sto fit results}). We therefore conclude that our new fitting methods provide reliable elastic moduli, even for samples with irregular geometries. 

\textbf{\mnge.} Next, we test \mnge, point group $D_{6h}$, which has 5 independent elastic moduli. The results of the fits are given in \autoref{table:mn3ge fit results}. Similar to \sto , we show fit results for a rectangular parallelepiped sample (sample A) with dimensions $\left( 0.87 \times 1.01 \times 1.19 \right)$~mm$^3$ and corners parallel to high-symmetry directions, as well as an irregularly-shaped sample (sample B---see \autoref{fig:sketch} for a 3D model).

As with \sto, all fit methods yield the same moduli for \mnge for both samples to within measurement uncertainty. Our results for sample A are also in agreement with previously-published elastic moduli of \mnge \cite{Theuss2022}. Comparing samples A and B, the absolute difference in elastic moduli stays below 4~GPa. This value is consistent with our results for \sto and is likely due to similar systematic uncertainties described above.

\begin{table}[]
	% \centering
	\begin{tabular}{ccccc}
		\hline
		\hline
		Sample                  & Fit Method & $c_{11}$ & $c_{12}$ & $c_{44}$ \\
		\hline
		\multirow{3}{*}{\sto A} & RPR        & 322    & 104    & 125  \\
		                        & FEM        & 322    & 104    & 125  \\
		                        & SMI        & 322    & 104    & 125  \\
		\hline
		\multirow{2}{*}{\sto B} & FEM        & 317    & 103    & 122  \\
		                        & SMI        & 317    & 103    & 122  \\
		\hline
		\multirow{1}{*}{Bell et al. \cite{bell_elastic_1963}} &        & 317.3    & 102.3    & 123.4  \\
		\hline
		\multirow{1}{*}{Poindexter et al. \cite{poindexter_elastic_1958}} &        & 348.2    & 100.6    & 119.0 \\
		\hline
		\hline
	\end{tabular}
	\caption{\textbf{Elastic moduli of \sto.} The elastic moduli in GPa for both \sto samples: the sample polished into the shape of a rectangular parallelepiped (Sample A), and the irregularly-shaped sample (Sample B). Uncertainties are approximately $\pm$2~GPa (due to the finite force applied to the sample by the RUS apparatus) and $\pm$0.4~GPa (from up to 1 degree misalignment between the crystal axes and the sample mesh). Literature values are provided for comparison. }
	\label{table:sto fit results}
\end{table}

\begin{table}[]
	% \centering
	\begin{tabular}{ccccccc}
		\hline
		\hline
		Sample                 & Fit Method  & $c_{11}$ & $c_{12}$ & $c_{13}$ & $c_{33}$ & $c_{44}$ \\
		\hline
		\multirow{3}{*}{\mnge A} & RPR       & 130    & 44     & 13     & 202    & 48 \\
		                         & FEM       & 130    & 44     & 13	    & 202    & 48 \\
								 & SMI       & 130    & 44     & 13     & 202    & 48 \\
		\hline
		\multirow{2}{*}{\mnge B} & FEM       & 127    & 40     & 14     & 203    & 49 \\
								 & SMI       & 127    & 40     & 14	    & 203    & 49 \\
		\hline
		\hline
	\end{tabular}
	\caption{\textbf{Elastic moduli of \mnge.} The elastic moduli in GPa for both \mnge samples: a rectangular parallelepiped (sample A) and the irregularly-shaped sample (sample B). Uncertainties are approximately $\pm$2~GPa (due to the finite force applied to the sample by the RUS apparatus) and $\pm$2~GPa (from up to 1 degree misalignment between the crystal axes and the sample mesh).}
	\label{table:mn3ge fit results}
\end{table}

\section{Application to UT$\rm{\bf e}_2$}

Having demonstrated the successful extraction of elastic moduli from RUS spectra of irregular samples, we now fit the elastic moduli for an irregularly-shaped sample of \ute. A 3D model of the sample is shown in \autoref{fig:sketch}, and the moduli are given in \autoref{table:ute2 fit results}.
\begin{table}%[hbt!]
	% \centering
	\begin{tabular}{ccccccccccc}
		\hline
		\hline
		 &	$T$ (K)    & $c_{11}$ & $c_{22}$ & $c_{33}$ & $c_{12}$ & $c_{13}$ & $c_{23}$ & $c_{44}$ & $c_{55}$ & $c_{66}$ \\
		\hline
		\multirow{2}{*}{RUS} &     4         & 90.3     & 144.1    &  95.9    & 25.7     & 41.3     & 31.9     & 28.0     &  53.2    &  30.4    \\
		   &	 300       & 84.7     & 139.5    &  91.1    & 26.8     & 38.1     & 31.6     & 26.9     &  52.4    &  29.7    \\
		\hline
		DFT    &                 & 96     & 136    & 90    & 28     & 46    & 26     & 28    & 57     & 31    \\ 
		\hline
		\hline
	\end{tabular}
	\caption{\textbf{Elastic moduli of \ute.} The elastic moduli of the \ute sample shown in \autoref{fig:sketch} at 300~K and at 4~K. Uncertainties are approximately $\pm$0.5~GPa (due to the finite force applied to the sample by the RUS apparatus) and $\pm$0.5~GPa (from a 1 degree misalignment between the crystal axes and the sample mesh). Moduli obtained from DFT calculations, with atomic coordinates optimized, are given on the bottom row. These values were used as rough guides to constrain the parameter space of the genetic algorithm fits to the RUS data.}
	\label{table:ute2 fit results}
\end{table}
% \begin{table}%[hbt!]
% 	% \centering
% 	\begin{tabular}{cccc}
% 		\hline
% 		\hline
% 		Elastic     & \multicolumn{2}{c}{RUS}  & DFT \\
% 					\cline{2-3}
% 		Modulus     & 4~K          & 300~K     &     \\
% 		\hline
% 		$c_{11}$    & 90.3         & 84.7      & 96  \\
% 		$c_{22}$    & 144.1        & 139.5     & 136 \\
% 		$c_{33}$    & 95.9         & 91.1      & 90  \\
% 		$c_{12}$    & 25.7         & 26.8      & 28  \\
% 		$c_{13}$    & 41.3         & 38.1      & 46  \\
% 		$c_{23}$    & 31.9         & 31.6      & 26  \\
% 		$c_{44}$    & 28.0         & 26.9      & 28  \\
% 		$c_{55}$    & 53.2         & 52.4      & 57  \\
% 		$c_{66}$    & 30.4         & 29.7      & 31  \\               
% 		\hline
% 		\hline
% 	\end{tabular}
% 	\caption{\textbf{Elastic moduli of \ute.} The elastic moduli in GPa of the \ute sample shown in \autoref{fig:sketch} at 300~K and at 4~K. Uncertainties are approximately $\pm$0.5~GPa (due to the finite force applied to the sample by the RUS apparatus) and $\pm$0.5~GPa (from a 1 degree misalignment between the crystal axes and the sample mesh). Moduli obtained from DFT calculations, with atomic coordinates optimized, are given on the bottom row. These values were used as rough guides to constrain the parameter space of the genetic algorithm fits to the RUS data.}
% 	\label{table:ute2 fit results}
% \end{table}

\ute is orthorhombic (point group $D_{2h}$), with nine independent elastic moduli. The population size---the number of initial guesses---required for a good fit using a genetic algorithm scales roughly linearly with the number of free parameters \cite{storn1997DifferentialEvolutionSimple}. This means that the \ute fits require nearly twice the population as compared to the previous \mnge fits. Additionally, the samples of \sto and \mnge were produced by adding additional facets to what were previously rectangular paraellelpiped samples, whereas our \ute sample is as-grown. This results in smaller feature sizes on the \ute sample as compared to the previous samples, requiring a finer mesh size for the 3D models (the full meshes are shown in the supplement). 

Both the dense mesh of our particular sample, and large number of moduli for \ute in general, increase the fit time for the FEM method, which must re-compute the entire spectrum at each stage of the fit. As noted in \citet{shragai2023RapidMethodComputing}, this makes FEM unsuitable for fitting under these circumstances, as convergence would take upward of a month. The SMI method, on the other hand, is largely unaffected by the increase in population size and mesh density because the computationally-intensive step is performed only once, at the start of the fit, and the results are stored for use in subsequent generations of the genetic algorithm. We therefore only perform SMI fits to our \ute spectra. This approach is justified by the results of the previous section, which demonstrated that fits using both FEM and SMI methods produce the same elastic moduli.

The elastic moduli of \ute, at both 300 K and 4 K are given in \autoref{table:ute2 fit results}. For comparison, we also report elastic moduli obtained from density-functional theory calculations (bottom row in \autoref{table:ute2 fit results}). These calculated values are in remarkable agreement with the experimental results. Other experimentally-relevant quantities, including the bulk modulus, the Young's moduli, and the Poisson's ratios, are tabulated in the supplement.

%We observe, that among all compressional elastic moduli in \ute, $c_{22}$, which corresponds to $\varepsilon_{yy}$ strain, is by far the largest. $c_{11}$ and $c_{33}$, on the other hand are quite similar in absolute value. This trend is reversed for the shear moduli, where $c_{55}$ \flo{(corresponding to $\varepsilon_{xz}$ strain)} is largest, and $c_{44}$ and $c_{66}$ have similar, but smaller values. \flo{These values show that while \ute reacts stiffest for compressions along the b axis, its shear response is most rigid in the perpendicular ac plane.} \st{This is a surprising observation, since $c_{55}$ corresponds to $\varepsilon_{xz}$, shear strain in the xz plane, which is perpendicular to $\varepsilon_{yy}$, the strain related to $c_{22}$.}
%\comments{\st{Crystal structure of ute2 has uranium ladders in xz plane. Maybe those don't like to be deformed? These ladders don't like to be pushed together? Also there are tellurium chains along y axis. Maybe superexchange between U atoms (more sensitive to angles) but direct exchange between Te atoms (more sensitive to distance)? These considerations might be stuff we can infer from the paragraph above, but is this way too much bs here? Or do you think it's worth spending some time thinking about it?}} \brad{(Yeah we should have a conversation about this paragraph and whether we need it here.)} \comments{I guess we talked about it briefly and said we should cut this paragraph. Does that mean just cut the "interpretation part" or also the two sentences before where we just make the observation?}

\section{Conclusion}
We implement two new methods for performing the forward calculation for resonance spectra in a genetic algorithm. These methods---SMI and FEM---allow us to fit the elastic moduli of irregularly-shaped samples using resonant ultrasound spectroscopy. This is first demonstrated on materials with known elastic moduli (\sto and \mnge), demonstrating consistency between our new methods and the older method of \citet{visscher_normal_1991}, which is only applicable for rectangular parallelepipeds. We then apply our new method to measure the full elastic tensor of the unconventional superconductor \ute at 300~K and at 4~K (\autoref{table:ute2 fit results}). 

We expect that the elastic moduli of \ute will be of use to researchers studying the superconducting properties under both uniaxial strain and hydrostatic pressure \cite{thomasSpatiallyInhomogeneousSuperconductivity2021,girod2022ThermodynamicElectricalTransport, thomas_evidence_2020, aokiMultipleSuperconductingPhases2020,braithwaiteMultipleSuperconductingPhases2019,ranEnhancementReentranceSpin2020}. We expect that the general method that we have introduced---implementing the fast surface mesh integration method into a genetic algorithm---will be broadly useful for measuring the elastic moduli of samples that cannot be easily prepared into regular geometric shapes.

\section{Acknowledgments}
B. J. R, F. T., G. G., and A. S. acknowledge funding from the Office of Basic Energy Sciences of the United States Department of Energy under Award No. DE-SC0020143 (UTe$_2$ and SrTiO$_3$ sample preparation, ultrasound experiments, and data analysis). B. J. R., F. T., G. G., A. S., T. C., and S. N. acknowledge funding from the Institute for Quantum Matter, an Energy Frontier Research Center funded by the Office of Basic Energy Sciences of the United States Department of Energy under Award No. DE-SC0019331 (Mn$_3$Ge sample growth and preparation). Imaging data was acquired through the Cornell Institute of Biotechnology's Imaging Facility, with NIH S10OD012287 for the ZEISS/Xradia Versa 520 X-ray Microscope (CT). Laue backreflection diffraction measurements made use of the Cornell Center for Materials Research Shared Facilities which are supported through the NSF MRSEC program (DMR-1719875). Research at the University of Maryland was supported by Department of Energy under award no. DE-SC-0019154 (sample characterization), the Gordon and Betty Moore Foundation’s EPiQS Initiative through grant no. GBMF9071 (materials synthesis), NIST, and the Maryland Quantum Materials Center. Discussion of commercial software does not imply endorsement by NIST.

\bibliography{literature}

%apsrev4-2.bst 2019-01-14 (MD) hand-edited version of apsrev4-1.bst
%Control: key (0)
%Control: author (8) initials jnrlst
%Control: editor formatted (1) identically to author
%Control: production of article title (0) allowed
%Control: page (0) single
%Control: year (1) truncated
%Control: production of eprint (0) enabled
\begin{thebibliography}{37}%
\makeatletter
\providecommand \@ifxundefined [1]{%
 \@ifx{#1\undefined}
}%
\providecommand \@ifnum [1]{%
 \ifnum #1\expandafter \@firstoftwo
 \else \expandafter \@secondoftwo
 \fi
}%
\providecommand \@ifx [1]{%
 \ifx #1\expandafter \@firstoftwo
 \else \expandafter \@secondoftwo
 \fi
}%
\providecommand \natexlab [1]{#1}%
\providecommand \enquote  [1]{``#1''}%
\providecommand \bibnamefont  [1]{#1}%
\providecommand \bibfnamefont [1]{#1}%
\providecommand \citenamefont [1]{#1}%
\providecommand \href@noop [0]{\@secondoftwo}%
\providecommand \href [0]{\begingroup \@sanitize@url \@href}%
\providecommand \@href[1]{\@@startlink{#1}\@@href}%
\providecommand \@@href[1]{\endgroup#1\@@endlink}%
\providecommand \@sanitize@url [0]{\catcode `\\12\catcode `\$12\catcode
  `\&12\catcode `\#12\catcode `\^12\catcode `\_12\catcode `\%12\relax}%
\providecommand \@@startlink[1]{}%
\providecommand \@@endlink[0]{}%
\providecommand \url  [0]{\begingroup\@sanitize@url \@url }%
\providecommand \@url [1]{\endgroup\@href {#1}{\urlprefix }}%
\providecommand \urlprefix  [0]{URL }%
\providecommand \Eprint [0]{\href }%
\providecommand \doibase [0]{https://doi.org/}%
\providecommand \selectlanguage [0]{\@gobble}%
\providecommand \bibinfo  [0]{\@secondoftwo}%
\providecommand \bibfield  [0]{\@secondoftwo}%
\providecommand \translation [1]{[#1]}%
\providecommand \BibitemOpen [0]{}%
\providecommand \bibitemStop [0]{}%
\providecommand \bibitemNoStop [0]{.\EOS\space}%
\providecommand \EOS [0]{\spacefactor3000\relax}%
\providecommand \BibitemShut  [1]{\csname bibitem#1\endcsname}%
\let\auto@bib@innerbib\@empty
%</preamble>
\bibitem [{\citenamefont {Visscher}\ \emph {et~al.}(1991)\citenamefont
  {Visscher}, \citenamefont {Migliori}, \citenamefont {Bell},\ and\
  \citenamefont {Reinert}}]{visscher_normal_1991}%
  \BibitemOpen
  \bibfield  {author} {\bibinfo {author} {\bibfnamefont {W.~M.}\ \bibnamefont
  {Visscher}}, \bibinfo {author} {\bibfnamefont {A.}~\bibnamefont {Migliori}},
  \bibinfo {author} {\bibfnamefont {T.~M.}\ \bibnamefont {Bell}},\ and\
  \bibinfo {author} {\bibfnamefont {R.~A.}\ \bibnamefont {Reinert}},\
  }\bibfield  {title} {\bibinfo {title} {On the normal modes of free vibration
  of inhomogeneous and anisotropic elastic objects},\ }\href
  {https://doi.org/10.1121/1.401643} {\bibfield  {journal} {\bibinfo  {journal}
  {The Journal of the Acoustical Society of America}\ }\textbf {\bibinfo
  {volume} {90}},\ \bibinfo {pages} {2154} (\bibinfo {year}
  {1991})}\BibitemShut {NoStop}%
\bibitem [{\citenamefont {Ran}\ \emph {et~al.}(2019{\natexlab{a}})\citenamefont
  {Ran}, \citenamefont {Eckberg}, \citenamefont {Ding}, \citenamefont
  {Furukawa}, \citenamefont {Metz}, \citenamefont {Saha}, \citenamefont {Liu},
  \citenamefont {Zic}, \citenamefont {Kim}, \citenamefont {Paglione},\ and\
  \citenamefont {Butch}}]{ranNearlyFerromagneticSpintriplet2019}%
  \BibitemOpen
  \bibfield  {author} {\bibinfo {author} {\bibfnamefont {S.}~\bibnamefont
  {Ran}}, \bibinfo {author} {\bibfnamefont {C.}~\bibnamefont {Eckberg}},
  \bibinfo {author} {\bibfnamefont {Q.-P.}\ \bibnamefont {Ding}}, \bibinfo
  {author} {\bibfnamefont {Y.}~\bibnamefont {Furukawa}}, \bibinfo {author}
  {\bibfnamefont {T.}~\bibnamefont {Metz}}, \bibinfo {author} {\bibfnamefont
  {S.~R.}\ \bibnamefont {Saha}}, \bibinfo {author} {\bibfnamefont {I.-L.}\
  \bibnamefont {Liu}}, \bibinfo {author} {\bibfnamefont {M.}~\bibnamefont
  {Zic}}, \bibinfo {author} {\bibfnamefont {H.}~\bibnamefont {Kim}}, \bibinfo
  {author} {\bibfnamefont {J.}~\bibnamefont {Paglione}},\ and\ \bibinfo
  {author} {\bibfnamefont {N.~P.}\ \bibnamefont {Butch}},\ }\bibfield  {title}
  {\bibinfo {title} {Nearly ferromagnetic spin-triplet superconductivity},\
  }\href {https://doi.org/10.1126/science.aav8645} {\bibfield  {journal}
  {\bibinfo  {journal} {Science}\ }\textbf {\bibinfo {volume} {365}},\ \bibinfo
  {pages} {684} (\bibinfo {year} {2019}{\natexlab{a}})}\BibitemShut {NoStop}%
\bibitem [{\citenamefont {Aoki}\ \emph {et~al.}(2019)\citenamefont {Aoki},
  \citenamefont {Nakamura}, \citenamefont {Honda}, \citenamefont {Li},
  \citenamefont {Homma}, \citenamefont {Shimizu}, \citenamefont {Sato},
  \citenamefont {Knebel}, \citenamefont {Brison}, \citenamefont {Pourret},
  \citenamefont {Braithwaite}, \citenamefont {Lapertot}, \citenamefont {Niu},
  \citenamefont {Vali{\v s}ka}, \citenamefont {Harima},\ and\ \citenamefont
  {Flouquet}}]{aokiUnconventionalSuperconductivityHeavy2019}%
  \BibitemOpen
  \bibfield  {author} {\bibinfo {author} {\bibfnamefont {D.}~\bibnamefont
  {Aoki}}, \bibinfo {author} {\bibfnamefont {A.}~\bibnamefont {Nakamura}},
  \bibinfo {author} {\bibfnamefont {F.}~\bibnamefont {Honda}}, \bibinfo
  {author} {\bibfnamefont {D.}~\bibnamefont {Li}}, \bibinfo {author}
  {\bibfnamefont {Y.}~\bibnamefont {Homma}}, \bibinfo {author} {\bibfnamefont
  {Y.}~\bibnamefont {Shimizu}}, \bibinfo {author} {\bibfnamefont {Y.~J.}\
  \bibnamefont {Sato}}, \bibinfo {author} {\bibfnamefont {G.}~\bibnamefont
  {Knebel}}, \bibinfo {author} {\bibfnamefont {J.-P.}\ \bibnamefont {Brison}},
  \bibinfo {author} {\bibfnamefont {A.}~\bibnamefont {Pourret}}, \bibinfo
  {author} {\bibfnamefont {D.}~\bibnamefont {Braithwaite}}, \bibinfo {author}
  {\bibfnamefont {G.}~\bibnamefont {Lapertot}}, \bibinfo {author}
  {\bibfnamefont {Q.}~\bibnamefont {Niu}}, \bibinfo {author} {\bibfnamefont
  {M.}~\bibnamefont {Vali{\v s}ka}}, \bibinfo {author} {\bibfnamefont
  {H.}~\bibnamefont {Harima}},\ and\ \bibinfo {author} {\bibfnamefont
  {J.}~\bibnamefont {Flouquet}},\ }\bibfield  {title} {\bibinfo {title}
  {{Unconventional Superconductivity in Heavy Fermion UTe$_2$}},\ }\href
  {https://doi.org/10.7566/JPSJ.88.043702} {\bibfield  {journal} {\bibinfo
  {journal} {Journal of the Physical Society of Japan}\ }\textbf {\bibinfo
  {volume} {88}},\ \bibinfo {pages} {043702} (\bibinfo {year}
  {2019})}\BibitemShut {NoStop}%
\bibitem [{\citenamefont {Nakamine}\ \emph {et~al.}(2019)\citenamefont
  {Nakamine}, \citenamefont {Kitagawa}, \citenamefont {Ishida}, \citenamefont
  {Tokunaga}, \citenamefont {Sakai}, \citenamefont {Kambe}, \citenamefont
  {Nakamura}, \citenamefont {Shimizu}, \citenamefont {Homma}, \citenamefont
  {Li}, \citenamefont {Honda},\ and\ \citenamefont
  {Aoki}}]{nakamineSuperconductingPropertiesHeavy2019}%
  \BibitemOpen
  \bibfield  {author} {\bibinfo {author} {\bibfnamefont {G.}~\bibnamefont
  {Nakamine}}, \bibinfo {author} {\bibfnamefont {S.}~\bibnamefont {Kitagawa}},
  \bibinfo {author} {\bibfnamefont {K.}~\bibnamefont {Ishida}}, \bibinfo
  {author} {\bibfnamefont {Y.}~\bibnamefont {Tokunaga}}, \bibinfo {author}
  {\bibfnamefont {H.}~\bibnamefont {Sakai}}, \bibinfo {author} {\bibfnamefont
  {S.}~\bibnamefont {Kambe}}, \bibinfo {author} {\bibfnamefont
  {A.}~\bibnamefont {Nakamura}}, \bibinfo {author} {\bibfnamefont
  {Y.}~\bibnamefont {Shimizu}}, \bibinfo {author} {\bibfnamefont
  {Y.}~\bibnamefont {Homma}}, \bibinfo {author} {\bibfnamefont
  {D.}~\bibnamefont {Li}}, \bibinfo {author} {\bibfnamefont {F.}~\bibnamefont
  {Honda}},\ and\ \bibinfo {author} {\bibfnamefont {D.}~\bibnamefont {Aoki}},\
  }\bibfield  {title} {\bibinfo {title} {Superconducting {{Properties}} of
  {{Heavy Fermion UTe$_2$}} {{Revealed}} by {{$^{125}$Te-nuclear Magnetic
  Resonance}}},\ }\href {https://doi.org/10.7566/JPSJ.88.113703} {\bibfield
  {journal} {\bibinfo  {journal} {Journal of the Physical Society of Japan}\
  }\textbf {\bibinfo {volume} {88}},\ \bibinfo {pages} {113703} (\bibinfo
  {year} {2019})}\BibitemShut {NoStop}%
\bibitem [{\citenamefont {Nakamine}\ \emph {et~al.}(2021)\citenamefont
  {Nakamine}, \citenamefont {Kinjo}, \citenamefont {Kitagawa}, \citenamefont
  {Ishida}, \citenamefont {Tokunaga}, \citenamefont {Sakai}, \citenamefont
  {Kambe}, \citenamefont {Nakamura}, \citenamefont {Shimizu}, \citenamefont
  {Homma}, \citenamefont {Li}, \citenamefont {Honda},\ and\ \citenamefont
  {Aoki}}]{nakamineInhomogeneousSuperconductingState2021}%
  \BibitemOpen
  \bibfield  {author} {\bibinfo {author} {\bibfnamefont {G.}~\bibnamefont
  {Nakamine}}, \bibinfo {author} {\bibfnamefont {K.}~\bibnamefont {Kinjo}},
  \bibinfo {author} {\bibfnamefont {S.}~\bibnamefont {Kitagawa}}, \bibinfo
  {author} {\bibfnamefont {K.}~\bibnamefont {Ishida}}, \bibinfo {author}
  {\bibfnamefont {Y.}~\bibnamefont {Tokunaga}}, \bibinfo {author}
  {\bibfnamefont {H.}~\bibnamefont {Sakai}}, \bibinfo {author} {\bibfnamefont
  {S.}~\bibnamefont {Kambe}}, \bibinfo {author} {\bibfnamefont
  {A.}~\bibnamefont {Nakamura}}, \bibinfo {author} {\bibfnamefont
  {Y.}~\bibnamefont {Shimizu}}, \bibinfo {author} {\bibfnamefont
  {Y.}~\bibnamefont {Homma}}, \bibinfo {author} {\bibfnamefont
  {D.}~\bibnamefont {Li}}, \bibinfo {author} {\bibfnamefont {F.}~\bibnamefont
  {Honda}},\ and\ \bibinfo {author} {\bibfnamefont {D.}~\bibnamefont {Aoki}},\
  }\bibfield  {title} {\bibinfo {title} {{Inhomogeneous Superconducting State
  Probed by $^{125}$Te NMR on UTe$_2$}},\ }\href
  {https://doi.org/10.7566/JPSJ.90.064709} {\bibfield  {journal} {\bibinfo
  {journal} {Journal of the Physical Society of Japan}\ }\textbf {\bibinfo
  {volume} {90}},\ \bibinfo {pages} {064709} (\bibinfo {year}
  {2021})}\BibitemShut {NoStop}%
\bibitem [{\citenamefont {Fujibayashi}\ \emph {et~al.}(2022)\citenamefont
  {Fujibayashi}, \citenamefont {Nakamine}, \citenamefont {Kinjo}, \citenamefont
  {Kitagawa}, \citenamefont {Ishida}, \citenamefont {Tokunaga}, \citenamefont
  {Sakai}, \citenamefont {Kambe}, \citenamefont {Nakamura}, \citenamefont
  {Shimizu}, \citenamefont {Homma}, \citenamefont {Li}, \citenamefont {Honda},\
  and\ \citenamefont {Aoki}}]{fujibayashiSuperconductingOrderParameter2022}%
  \BibitemOpen
  \bibfield  {author} {\bibinfo {author} {\bibfnamefont {H.}~\bibnamefont
  {Fujibayashi}}, \bibinfo {author} {\bibfnamefont {G.}~\bibnamefont
  {Nakamine}}, \bibinfo {author} {\bibfnamefont {K.}~\bibnamefont {Kinjo}},
  \bibinfo {author} {\bibfnamefont {S.}~\bibnamefont {Kitagawa}}, \bibinfo
  {author} {\bibfnamefont {K.}~\bibnamefont {Ishida}}, \bibinfo {author}
  {\bibfnamefont {Y.}~\bibnamefont {Tokunaga}}, \bibinfo {author}
  {\bibfnamefont {H.}~\bibnamefont {Sakai}}, \bibinfo {author} {\bibfnamefont
  {S.}~\bibnamefont {Kambe}}, \bibinfo {author} {\bibfnamefont
  {A.}~\bibnamefont {Nakamura}}, \bibinfo {author} {\bibfnamefont
  {Y.}~\bibnamefont {Shimizu}}, \bibinfo {author} {\bibfnamefont
  {Y.}~\bibnamefont {Homma}}, \bibinfo {author} {\bibfnamefont
  {D.}~\bibnamefont {Li}}, \bibinfo {author} {\bibfnamefont {F.}~\bibnamefont
  {Honda}},\ and\ \bibinfo {author} {\bibfnamefont {D.}~\bibnamefont {Aoki}},\
  }\bibfield  {title} {\bibinfo {title} {{Superconducting Order Parameter in
  UTe$_2$ Determined by Knight Shift Measurement}},\ }\href
  {https://doi.org/10.7566/JPSJ.91.043705} {\bibfield  {journal} {\bibinfo
  {journal} {Journal of the Physical Society of Japan}\ }\textbf {\bibinfo
  {volume} {91}},\ \bibinfo {pages} {043705} (\bibinfo {year}
  {2022})}\BibitemShut {NoStop}%
\bibitem [{\citenamefont {Knebel}\ \emph {et~al.}(2019)\citenamefont {Knebel},
  \citenamefont {Knafo}, \citenamefont {Pourret}, \citenamefont {Niu},
  \citenamefont {Vali{\v s}ka}, \citenamefont {Braithwaite}, \citenamefont
  {Lapertot}, \citenamefont {Nardone}, \citenamefont {Zitouni}, \citenamefont
  {Mishra}, \citenamefont {Sheikin}, \citenamefont {Seyfarth}, \citenamefont
  {Brison}, \citenamefont {Aoki},\ and\ \citenamefont
  {Flouquet}}]{knebelFieldreentrantSuperconductivityClose2019}%
  \BibitemOpen
  \bibfield  {author} {\bibinfo {author} {\bibfnamefont {G.}~\bibnamefont
  {Knebel}}, \bibinfo {author} {\bibfnamefont {W.}~\bibnamefont {Knafo}},
  \bibinfo {author} {\bibfnamefont {A.}~\bibnamefont {Pourret}}, \bibinfo
  {author} {\bibfnamefont {Q.}~\bibnamefont {Niu}}, \bibinfo {author}
  {\bibfnamefont {M.}~\bibnamefont {Vali{\v s}ka}}, \bibinfo {author}
  {\bibfnamefont {D.}~\bibnamefont {Braithwaite}}, \bibinfo {author}
  {\bibfnamefont {G.}~\bibnamefont {Lapertot}}, \bibinfo {author}
  {\bibfnamefont {M.}~\bibnamefont {Nardone}}, \bibinfo {author} {\bibfnamefont
  {A.}~\bibnamefont {Zitouni}}, \bibinfo {author} {\bibfnamefont
  {S.}~\bibnamefont {Mishra}}, \bibinfo {author} {\bibfnamefont
  {I.}~\bibnamefont {Sheikin}}, \bibinfo {author} {\bibfnamefont
  {G.}~\bibnamefont {Seyfarth}}, \bibinfo {author} {\bibfnamefont {J.~P.}\
  \bibnamefont {Brison}}, \bibinfo {author} {\bibfnamefont {D.}~\bibnamefont
  {Aoki}},\ and\ \bibinfo {author} {\bibfnamefont {J.}~\bibnamefont
  {Flouquet}},\ }\bibfield  {title} {\bibinfo {title} {{Field-Reentrant
  Superconductivity Close to a Metamagnetic Transition in the Heavy-Fermion
  Superconductor UTe$_2$}},\ }\href {https://doi.org/10.7566/JPSJ.88.063707}
  {\bibfield  {journal} {\bibinfo  {journal} {Journal of the Physical Society
  of Japan}\ }\textbf {\bibinfo {volume} {88}},\ \bibinfo {pages} {063707}
  (\bibinfo {year} {2019})}\BibitemShut {NoStop}%
\bibitem [{\citenamefont {Ran}\ \emph {et~al.}(2019{\natexlab{b}})\citenamefont
  {Ran}, \citenamefont {Liu}, \citenamefont {Eo}, \citenamefont {Campbell},
  \citenamefont {Neves}, \citenamefont {Fuhrman}, \citenamefont {Saha},
  \citenamefont {Eckberg}, \citenamefont {Kim}, \citenamefont {Graf},
  \citenamefont {Balakirev}, \citenamefont {Singleton}, \citenamefont
  {Paglione},\ and\ \citenamefont
  {Butch}}]{ranExtremeMagneticFieldboosted2019}%
  \BibitemOpen
  \bibfield  {author} {\bibinfo {author} {\bibfnamefont {S.}~\bibnamefont
  {Ran}}, \bibinfo {author} {\bibfnamefont {I.~L.}\ \bibnamefont {Liu}},
  \bibinfo {author} {\bibfnamefont {Y.~S.}\ \bibnamefont {Eo}}, \bibinfo
  {author} {\bibfnamefont {D.~J.}\ \bibnamefont {Campbell}}, \bibinfo {author}
  {\bibfnamefont {P.~M.}\ \bibnamefont {Neves}}, \bibinfo {author}
  {\bibfnamefont {W.~T.}\ \bibnamefont {Fuhrman}}, \bibinfo {author}
  {\bibfnamefont {S.~R.}\ \bibnamefont {Saha}}, \bibinfo {author}
  {\bibfnamefont {C.}~\bibnamefont {Eckberg}}, \bibinfo {author} {\bibfnamefont
  {H.}~\bibnamefont {Kim}}, \bibinfo {author} {\bibfnamefont {D.}~\bibnamefont
  {Graf}}, \bibinfo {author} {\bibfnamefont {F.}~\bibnamefont {Balakirev}},
  \bibinfo {author} {\bibfnamefont {J.}~\bibnamefont {Singleton}}, \bibinfo
  {author} {\bibfnamefont {J.}~\bibnamefont {Paglione}},\ and\ \bibinfo
  {author} {\bibfnamefont {N.~P.}\ \bibnamefont {Butch}},\ }\bibfield  {title}
  {\bibinfo {title} {Extreme magnetic field-boosted superconductivity},\ }\href
  {https://doi.org/10.1038/s41567-019-0670-x} {\bibfield  {journal} {\bibinfo
  {journal} {Nature Physics}\ }\textbf {\bibinfo {volume} {15}},\ \bibinfo
  {pages} {1250} (\bibinfo {year} {2019}{\natexlab{b}})}\BibitemShut {NoStop}%
\bibitem [{\citenamefont {Ran}\ \emph {et~al.}(2020)\citenamefont {Ran},
  \citenamefont {Kim}, \citenamefont {Liu}, \citenamefont {Saha}, \citenamefont
  {Hayes}, \citenamefont {Metz}, \citenamefont {Eo}, \citenamefont {Paglione},\
  and\ \citenamefont {Butch}}]{ranEnhancementReentranceSpin2020}%
  \BibitemOpen
  \bibfield  {author} {\bibinfo {author} {\bibfnamefont {S.}~\bibnamefont
  {Ran}}, \bibinfo {author} {\bibfnamefont {H.}~\bibnamefont {Kim}}, \bibinfo
  {author} {\bibfnamefont {I.-L.}\ \bibnamefont {Liu}}, \bibinfo {author}
  {\bibfnamefont {S.~R.}\ \bibnamefont {Saha}}, \bibinfo {author}
  {\bibfnamefont {I.}~\bibnamefont {Hayes}}, \bibinfo {author} {\bibfnamefont
  {T.}~\bibnamefont {Metz}}, \bibinfo {author} {\bibfnamefont {Y.~S.}\
  \bibnamefont {Eo}}, \bibinfo {author} {\bibfnamefont {J.}~\bibnamefont
  {Paglione}},\ and\ \bibinfo {author} {\bibfnamefont {N.~P.}\ \bibnamefont
  {Butch}},\ }\bibfield  {title} {\bibinfo {title} {Enhancement and reentrance
  of spin triplet superconductivity in {UTe$_2$} under pressure},\ }\href
  {https://doi.org/10.1103/PhysRevB.101.140503} {\bibfield  {journal} {\bibinfo
   {journal} {Physical Review B}\ }\textbf {\bibinfo {volume} {101}},\ \bibinfo
  {pages} {140503} (\bibinfo {year} {2020})}\BibitemShut {NoStop}%
\bibitem [{\citenamefont {Aoki}\ \emph {et~al.}(2020)\citenamefont {Aoki},
  \citenamefont {Honda}, \citenamefont {Knebel}, \citenamefont {Braithwaite},
  \citenamefont {Nakamura}, \citenamefont {Li}, \citenamefont {Homma},
  \citenamefont {Shimizu}, \citenamefont {Sato}, \citenamefont {Brison},\ and\
  \citenamefont {Flouquet}}]{aokiMultipleSuperconductingPhases2020}%
  \BibitemOpen
  \bibfield  {author} {\bibinfo {author} {\bibfnamefont {D.}~\bibnamefont
  {Aoki}}, \bibinfo {author} {\bibfnamefont {F.}~\bibnamefont {Honda}},
  \bibinfo {author} {\bibfnamefont {G.}~\bibnamefont {Knebel}}, \bibinfo
  {author} {\bibfnamefont {D.}~\bibnamefont {Braithwaite}}, \bibinfo {author}
  {\bibfnamefont {A.}~\bibnamefont {Nakamura}}, \bibinfo {author}
  {\bibfnamefont {D.~X.}\ \bibnamefont {Li}}, \bibinfo {author} {\bibfnamefont
  {Y.}~\bibnamefont {Homma}}, \bibinfo {author} {\bibfnamefont
  {Y.}~\bibnamefont {Shimizu}}, \bibinfo {author} {\bibfnamefont {Y.~J.}\
  \bibnamefont {Sato}}, \bibinfo {author} {\bibfnamefont {J.~P.}\ \bibnamefont
  {Brison}},\ and\ \bibinfo {author} {\bibfnamefont {J.}~\bibnamefont
  {Flouquet}},\ }\bibfield  {title} {\bibinfo {title} {{Multiple
  Superconducting Phases and Unusual Enhancement of the Upper Critical Field in
  UTe$_2$}},\ }\href {https://doi.org/10.7566/JPSJ.89.053705} {\bibfield
  {journal} {\bibinfo  {journal} {Journal of the Physical Society of Japan}\
  }\textbf {\bibinfo {volume} {89}},\ \bibinfo {pages} {053705} (\bibinfo
  {year} {2020})}\BibitemShut {NoStop}%
\bibitem [{\citenamefont {Wei}\ \emph {et~al.}(2022)\citenamefont {Wei},
  \citenamefont {Saykin}, \citenamefont {Miller}, \citenamefont {Ran},
  \citenamefont {Saha}, \citenamefont {Agterberg}, \citenamefont {Schmalian},
  \citenamefont {Butch}, \citenamefont {Paglione},\ and\ \citenamefont
  {Kapitulnik}}]{weiInterplayMagnetismSuperconductivity2021}%
  \BibitemOpen
  \bibfield  {author} {\bibinfo {author} {\bibfnamefont {D.~S.}\ \bibnamefont
  {Wei}}, \bibinfo {author} {\bibfnamefont {D.}~\bibnamefont {Saykin}},
  \bibinfo {author} {\bibfnamefont {O.~Y.}\ \bibnamefont {Miller}}, \bibinfo
  {author} {\bibfnamefont {S.}~\bibnamefont {Ran}}, \bibinfo {author}
  {\bibfnamefont {S.~R.}\ \bibnamefont {Saha}}, \bibinfo {author}
  {\bibfnamefont {D.~F.}\ \bibnamefont {Agterberg}}, \bibinfo {author}
  {\bibfnamefont {J.}~\bibnamefont {Schmalian}}, \bibinfo {author}
  {\bibfnamefont {N.~P.}\ \bibnamefont {Butch}}, \bibinfo {author}
  {\bibfnamefont {J.}~\bibnamefont {Paglione}},\ and\ \bibinfo {author}
  {\bibfnamefont {A.}~\bibnamefont {Kapitulnik}},\ }\bibfield  {title}
  {\bibinfo {title} {Interplay between magnetism and superconductivity in
  {UTe$_2$}},\ }\href {https://doi.org/10.1103/PhysRevB.105.024521} {\bibfield
  {journal} {\bibinfo  {journal} {Physical Review B}\ }\textbf {\bibinfo
  {volume} {105}},\ \bibinfo {pages} {024521} (\bibinfo {year}
  {2022})}\BibitemShut {NoStop}%
\bibitem [{\citenamefont {Hayes}\ \emph {et~al.}(2021)\citenamefont {Hayes},
  \citenamefont {Wei}, \citenamefont {Metz}, \citenamefont {Zhang},
  \citenamefont {Eo}, \citenamefont {Ran}, \citenamefont {Saha}, \citenamefont
  {Collini}, \citenamefont {Butch}, \citenamefont {Agterberg}, \citenamefont
  {Kapitulnik},\ and\ \citenamefont
  {Paglione}}]{hayesMulticomponentSuperconductingOrder2021}%
  \BibitemOpen
  \bibfield  {author} {\bibinfo {author} {\bibfnamefont {I.~M.}\ \bibnamefont
  {Hayes}}, \bibinfo {author} {\bibfnamefont {D.~S.}\ \bibnamefont {Wei}},
  \bibinfo {author} {\bibfnamefont {T.}~\bibnamefont {Metz}}, \bibinfo {author}
  {\bibfnamefont {J.}~\bibnamefont {Zhang}}, \bibinfo {author} {\bibfnamefont
  {Y.~S.}\ \bibnamefont {Eo}}, \bibinfo {author} {\bibfnamefont
  {S.}~\bibnamefont {Ran}}, \bibinfo {author} {\bibfnamefont {S.~R.}\
  \bibnamefont {Saha}}, \bibinfo {author} {\bibfnamefont {J.}~\bibnamefont
  {Collini}}, \bibinfo {author} {\bibfnamefont {N.~P.}\ \bibnamefont {Butch}},
  \bibinfo {author} {\bibfnamefont {D.~F.}\ \bibnamefont {Agterberg}}, \bibinfo
  {author} {\bibfnamefont {A.}~\bibnamefont {Kapitulnik}},\ and\ \bibinfo
  {author} {\bibfnamefont {J.}~\bibnamefont {Paglione}},\ }\bibfield  {title}
  {\bibinfo {title} {Multicomponent superconducting order parameter in
  {UTe$_2$}},\ }\href {https://doi.org/10.1126/science.abb0272} {\bibfield
  {journal} {\bibinfo  {journal} {Science}\ }\textbf {\bibinfo {volume}
  {373}},\ \bibinfo {pages} {797} (\bibinfo {year} {2021})}\BibitemShut
  {NoStop}%
\bibitem [{\citenamefont {Rosuel}\ \emph {et~al.}(2022)\citenamefont {Rosuel},
  \citenamefont {Marcenat}, \citenamefont {Knebel}, \citenamefont {Klein},
  \citenamefont {Pourret}, \citenamefont {Marquardt}, \citenamefont {Niu},
  \citenamefont {Rousseau}, \citenamefont {Demuer}, \citenamefont {Seyfarth},
  \citenamefont {Lapertot}, \citenamefont {Aoki}, \citenamefont {Braithwaite},
  \citenamefont {Flouquet},\ and\ \citenamefont
  {Brison}}]{rosuel2022FieldinducedTuningPairing}%
  \BibitemOpen
  \bibfield  {author} {\bibinfo {author} {\bibfnamefont {A.}~\bibnamefont
  {Rosuel}}, \bibinfo {author} {\bibfnamefont {C.}~\bibnamefont {Marcenat}},
  \bibinfo {author} {\bibfnamefont {G.}~\bibnamefont {Knebel}}, \bibinfo
  {author} {\bibfnamefont {T.}~\bibnamefont {Klein}}, \bibinfo {author}
  {\bibfnamefont {A.}~\bibnamefont {Pourret}}, \bibinfo {author} {\bibfnamefont
  {N.}~\bibnamefont {Marquardt}}, \bibinfo {author} {\bibfnamefont
  {Q.}~\bibnamefont {Niu}}, \bibinfo {author} {\bibfnamefont {S.}~\bibnamefont
  {Rousseau}}, \bibinfo {author} {\bibfnamefont {A.}~\bibnamefont {Demuer}},
  \bibinfo {author} {\bibfnamefont {G.}~\bibnamefont {Seyfarth}}, \bibinfo
  {author} {\bibfnamefont {G.}~\bibnamefont {Lapertot}}, \bibinfo {author}
  {\bibfnamefont {D.}~\bibnamefont {Aoki}}, \bibinfo {author} {\bibfnamefont
  {D.}~\bibnamefont {Braithwaite}}, \bibinfo {author} {\bibfnamefont
  {J.}~\bibnamefont {Flouquet}},\ and\ \bibinfo {author} {\bibfnamefont
  {J.-P.}\ \bibnamefont {Brison}},\ }\bibfield  {title} {\bibinfo {title}
  {Field-induced tuning of the pairing state in a superconductor},\ }\href
  {https://doi.org/10.48550/arXiv.2205.04524} {\bibfield  {journal} {\bibinfo
  {journal} {{arXiv}:2205.04524}\ } (\bibinfo {year} {2022})}\BibitemShut
  {NoStop}%
\bibitem [{\citenamefont {Cairns}\ \emph {et~al.}(2020)\citenamefont {Cairns},
  \citenamefont {Stevens}, \citenamefont {O'Neill},\ and\ \citenamefont
  {Huxley}}]{Cairns2020}%
  \BibitemOpen
  \bibfield  {author} {\bibinfo {author} {\bibfnamefont {L.~P.}\ \bibnamefont
  {Cairns}}, \bibinfo {author} {\bibfnamefont {C.~R.}\ \bibnamefont {Stevens}},
  \bibinfo {author} {\bibfnamefont {C.~D.}\ \bibnamefont {O'Neill}},\ and\
  \bibinfo {author} {\bibfnamefont {A.}~\bibnamefont {Huxley}},\ }\bibfield
  {title} {\bibinfo {title} {Composition dependence of the superconducting
  properties of {UTe$_2$}},\ }\href {https://doi.org/10.1088/1361-648X/ab9c5d}
  {\bibfield  {journal} {\bibinfo  {journal} {Journal of Physics Condensed
  Matter}\ }\textbf {\bibinfo {volume} {32}},\ \bibinfo {pages} {415602}
  (\bibinfo {year} {2020})}\BibitemShut {NoStop}%
\bibitem [{\citenamefont {Thomas}\ \emph {et~al.}(2021)\citenamefont {Thomas},
  \citenamefont {Stevens}, \citenamefont {Santos}, \citenamefont {Fender},
  \citenamefont {Bauer}, \citenamefont {Ronning}, \citenamefont {Thompson},
  \citenamefont {Huxley},\ and\ \citenamefont
  {Rosa}}]{thomasSpatiallyInhomogeneousSuperconductivity2021}%
  \BibitemOpen
  \bibfield  {author} {\bibinfo {author} {\bibfnamefont {S.~M.}\ \bibnamefont
  {Thomas}}, \bibinfo {author} {\bibfnamefont {C.}~\bibnamefont {Stevens}},
  \bibinfo {author} {\bibfnamefont {F.~B.}\ \bibnamefont {Santos}}, \bibinfo
  {author} {\bibfnamefont {S.~S.}\ \bibnamefont {Fender}}, \bibinfo {author}
  {\bibfnamefont {E.~D.}\ \bibnamefont {Bauer}}, \bibinfo {author}
  {\bibfnamefont {F.}~\bibnamefont {Ronning}}, \bibinfo {author} {\bibfnamefont
  {J.~D.}\ \bibnamefont {Thompson}}, \bibinfo {author} {\bibfnamefont
  {A.}~\bibnamefont {Huxley}},\ and\ \bibinfo {author} {\bibfnamefont
  {P.~F.~S.}\ \bibnamefont {Rosa}},\ }\bibfield  {title} {\bibinfo {title}
  {Spatially inhomogeneous superconductivity in {UTe$_2$}},\ }\href
  {https://doi.org/10.1103/PhysRevB.104.224501} {\bibfield  {journal} {\bibinfo
   {journal} {Physical Review B}\ }\textbf {\bibinfo {volume} {104}},\ \bibinfo
  {pages} {224501} (\bibinfo {year} {2021})}\BibitemShut {NoStop}%
\bibitem [{\citenamefont {Rosa}\ \emph {et~al.}(2022)\citenamefont {Rosa},
  \citenamefont {Weiland}, \citenamefont {Fender}, \citenamefont {Scott},
  \citenamefont {Ronning}, \citenamefont {Thompson}, \citenamefont {Bauer},\
  and\ \citenamefont {Thomas}}]{rosaSinglecomponentSuperconductingState2022}%
  \BibitemOpen
  \bibfield  {author} {\bibinfo {author} {\bibfnamefont {P.~F.~S.}\
  \bibnamefont {Rosa}}, \bibinfo {author} {\bibfnamefont {A.}~\bibnamefont
  {Weiland}}, \bibinfo {author} {\bibfnamefont {S.~S.}\ \bibnamefont {Fender}},
  \bibinfo {author} {\bibfnamefont {B.~L.}\ \bibnamefont {Scott}}, \bibinfo
  {author} {\bibfnamefont {F.}~\bibnamefont {Ronning}}, \bibinfo {author}
  {\bibfnamefont {J.~D.}\ \bibnamefont {Thompson}}, \bibinfo {author}
  {\bibfnamefont {E.~D.}\ \bibnamefont {Bauer}},\ and\ \bibinfo {author}
  {\bibfnamefont {S.~M.}\ \bibnamefont {Thomas}},\ }\bibfield  {title}
  {\bibinfo {title} {Single thermodynamic transition at 2 {K} in
  superconducting {UTe$_2$} single crystals},\ }\href
  {https://doi.org/10.1038/s43246-022-00254-2} {\bibfield  {journal} {\bibinfo
  {journal} {Communications Materials}\ }\textbf {\bibinfo {volume} {3}},\
  \bibinfo {pages} {33} (\bibinfo {year} {2022})}\BibitemShut {NoStop}%
\bibitem [{\citenamefont {Braithwaite}\ \emph {et~al.}(2019)\citenamefont
  {Braithwaite}, \citenamefont {Vali{\v s}ka}, \citenamefont {Knebel},
  \citenamefont {Lapertot}, \citenamefont {Brison}, \citenamefont {Pourret},
  \citenamefont {Zhitomirsky}, \citenamefont {Flouquet}, \citenamefont
  {Honda},\ and\ \citenamefont
  {Aoki}}]{braithwaiteMultipleSuperconductingPhases2019}%
  \BibitemOpen
  \bibfield  {author} {\bibinfo {author} {\bibfnamefont {D.}~\bibnamefont
  {Braithwaite}}, \bibinfo {author} {\bibfnamefont {M.}~\bibnamefont {Vali{\v
  s}ka}}, \bibinfo {author} {\bibfnamefont {G.}~\bibnamefont {Knebel}},
  \bibinfo {author} {\bibfnamefont {G.}~\bibnamefont {Lapertot}}, \bibinfo
  {author} {\bibfnamefont {J.-P.}\ \bibnamefont {Brison}}, \bibinfo {author}
  {\bibfnamefont {A.}~\bibnamefont {Pourret}}, \bibinfo {author} {\bibfnamefont
  {M.~E.}\ \bibnamefont {Zhitomirsky}}, \bibinfo {author} {\bibfnamefont
  {J.}~\bibnamefont {Flouquet}}, \bibinfo {author} {\bibfnamefont
  {F.}~\bibnamefont {Honda}},\ and\ \bibinfo {author} {\bibfnamefont
  {D.}~\bibnamefont {Aoki}},\ }\bibfield  {title} {\bibinfo {title} {Multiple
  superconducting phases in a nearly ferromagnetic system},\ }\href
  {https://doi.org/10.1038/s42005-019-0248-z} {\bibfield  {journal} {\bibinfo
  {journal} {Communications Physics}\ }\textbf {\bibinfo {volume} {2}},\
  \bibinfo {pages} {147} (\bibinfo {year} {2019})}\BibitemShut {NoStop}%
\bibitem [{\citenamefont {Girod}\ \emph {et~al.}(2022)\citenamefont {Girod},
  \citenamefont {Stevens}, \citenamefont {Huxley}, \citenamefont {Bauer},
  \citenamefont {Santos}, \citenamefont {Thompson}, \citenamefont {Fernandes},
  \citenamefont {Zhu}, \citenamefont {Ronning}, \citenamefont {Rosa},\ and\
  \citenamefont {Thomas}}]{girod2022ThermodynamicElectricalTransport}%
  \BibitemOpen
  \bibfield  {author} {\bibinfo {author} {\bibfnamefont {C.}~\bibnamefont
  {Girod}}, \bibinfo {author} {\bibfnamefont {C.~R.}\ \bibnamefont {Stevens}},
  \bibinfo {author} {\bibfnamefont {A.}~\bibnamefont {Huxley}}, \bibinfo
  {author} {\bibfnamefont {E.~D.}\ \bibnamefont {Bauer}}, \bibinfo {author}
  {\bibfnamefont {F.~B.}\ \bibnamefont {Santos}}, \bibinfo {author}
  {\bibfnamefont {J.~D.}\ \bibnamefont {Thompson}}, \bibinfo {author}
  {\bibfnamefont {R.~M.}\ \bibnamefont {Fernandes}}, \bibinfo {author}
  {\bibfnamefont {J.-X.}\ \bibnamefont {Zhu}}, \bibinfo {author} {\bibfnamefont
  {F.}~\bibnamefont {Ronning}}, \bibinfo {author} {\bibfnamefont {P.~F.~S.}\
  \bibnamefont {Rosa}},\ and\ \bibinfo {author} {\bibfnamefont {S.~M.}\
  \bibnamefont {Thomas}},\ }\bibfield  {title} {\bibinfo {title} {Thermodynamic
  and electrical transport properties of {UTe$_2$} under uniaxial stress},\
  }\href {https://doi.org/10.1103/PhysRevB.106.L121101} {\bibfield  {journal}
  {\bibinfo  {journal} {Physical Review B}\ }\textbf {\bibinfo {volume}
  {106}},\ \bibinfo {pages} {L121101} (\bibinfo {year} {2022})}\BibitemShut
  {NoStop}%
\bibitem [{\citenamefont {Barber}\ \emph {et~al.}(2019)\citenamefont {Barber},
  \citenamefont {Lechermann}, \citenamefont {Streltsov}, \citenamefont
  {Skornyakov}, \citenamefont {Ghosh}, \citenamefont {Ramshaw}, \citenamefont
  {Kikugawa}, \citenamefont {Sokolov}, \citenamefont {Mackenzie}, \citenamefont
  {Hicks} \emph {et~al.}}]{barber2019role}%
  \BibitemOpen
  \bibfield  {author} {\bibinfo {author} {\bibfnamefont {M.~E.}\ \bibnamefont
  {Barber}}, \bibinfo {author} {\bibfnamefont {F.}~\bibnamefont {Lechermann}},
  \bibinfo {author} {\bibfnamefont {S.~V.}\ \bibnamefont {Streltsov}}, \bibinfo
  {author} {\bibfnamefont {S.~L.}\ \bibnamefont {Skornyakov}}, \bibinfo
  {author} {\bibfnamefont {S.}~\bibnamefont {Ghosh}}, \bibinfo {author}
  {\bibfnamefont {B.}~\bibnamefont {Ramshaw}}, \bibinfo {author} {\bibfnamefont
  {N.}~\bibnamefont {Kikugawa}}, \bibinfo {author} {\bibfnamefont {D.~A.}\
  \bibnamefont {Sokolov}}, \bibinfo {author} {\bibfnamefont {A.~P.}\
  \bibnamefont {Mackenzie}}, \bibinfo {author} {\bibfnamefont {C.~W.}\
  \bibnamefont {Hicks}}, \emph {et~al.},\ }\bibfield  {title} {\bibinfo {title}
  {Role of correlations in determining the {Van Hove} strain in
  {Sr$_2$RuO$_4$}},\ }\href
  {https://doi.org/https://doi.org/10.1103/PhysRevB.100.245139} {\bibfield
  {journal} {\bibinfo  {journal} {Physical Review B}\ }\textbf {\bibinfo
  {volume} {100}},\ \bibinfo {pages} {245139} (\bibinfo {year}
  {2019})}\BibitemShut {NoStop}%
\bibitem [{\citenamefont {Li}\ \emph {et~al.}(2022)\citenamefont {Li},
  \citenamefont {Garst}, \citenamefont {Schmalian}, \citenamefont {Ghosh},
  \citenamefont {Kikugawa}, \citenamefont {Sokolov}, \citenamefont {Hicks},
  \citenamefont {Jerzembeck}, \citenamefont {Ikeda}, \citenamefont {Hu},
  \citenamefont {Ramshaw}, \citenamefont {Rost}, \citenamefont {Nicklas},\ and\
  \citenamefont {Mackenzie}}]{li2022elastocaloric}%
  \BibitemOpen
  \bibfield  {author} {\bibinfo {author} {\bibfnamefont {Y.-S.}\ \bibnamefont
  {Li}}, \bibinfo {author} {\bibfnamefont {M.}~\bibnamefont {Garst}}, \bibinfo
  {author} {\bibfnamefont {J.}~\bibnamefont {Schmalian}}, \bibinfo {author}
  {\bibfnamefont {S.}~\bibnamefont {Ghosh}}, \bibinfo {author} {\bibfnamefont
  {N.}~\bibnamefont {Kikugawa}}, \bibinfo {author} {\bibfnamefont {D.~A.}\
  \bibnamefont {Sokolov}}, \bibinfo {author} {\bibfnamefont {C.~W.}\
  \bibnamefont {Hicks}}, \bibinfo {author} {\bibfnamefont {F.}~\bibnamefont
  {Jerzembeck}}, \bibinfo {author} {\bibfnamefont {M.~S.}\ \bibnamefont
  {Ikeda}}, \bibinfo {author} {\bibfnamefont {Z.}~\bibnamefont {Hu}}, \bibinfo
  {author} {\bibfnamefont {B.~J.}\ \bibnamefont {Ramshaw}}, \bibinfo {author}
  {\bibfnamefont {A.~W.}\ \bibnamefont {Rost}}, \bibinfo {author}
  {\bibfnamefont {M.}~\bibnamefont {Nicklas}},\ and\ \bibinfo {author}
  {\bibfnamefont {A.~P.}\ \bibnamefont {Mackenzie}},\ }\bibfield  {title}
  {\bibinfo {title} {Elastocaloric determination of the phase diagram of
  {{Sr$_2$RuO$_4$}}},\ }\href {https://doi.org/10.1038/s41586-022-04820-z}
  {\bibfield  {journal} {\bibinfo  {journal} {Nature}\ }\textbf {\bibinfo
  {volume} {607}},\ \bibinfo {pages} {276} (\bibinfo {year}
  {2022})}\BibitemShut {NoStop}%
\bibitem [{\citenamefont {Bachmann}\ \emph {et~al.}(2019)\citenamefont
  {Bachmann}, \citenamefont {Ferguson}, \citenamefont {Theuss}, \citenamefont
  {Meng}, \citenamefont {Putzke}, \citenamefont {Helm}, \citenamefont {Shirer},
  \citenamefont {Li}, \citenamefont {Modic}, \citenamefont {Nicklas},
  \citenamefont {König}, \citenamefont {Low}, \citenamefont {Ghosh},
  \citenamefont {Mackenzie}, \citenamefont {Arnold}, \citenamefont {Hassinger},
  \citenamefont {{McDonald}}, \citenamefont {Winter}, \citenamefont {Bauer},
  \citenamefont {Ronning}, \citenamefont {Ramshaw}, \citenamefont {Nowack},\
  and\ \citenamefont {Moll}}]{bachmann2019spatial}%
  \BibitemOpen
  \bibfield  {author} {\bibinfo {author} {\bibfnamefont {M.~D.}\ \bibnamefont
  {Bachmann}}, \bibinfo {author} {\bibfnamefont {G.~M.}\ \bibnamefont
  {Ferguson}}, \bibinfo {author} {\bibfnamefont {F.}~\bibnamefont {Theuss}},
  \bibinfo {author} {\bibfnamefont {T.}~\bibnamefont {Meng}}, \bibinfo {author}
  {\bibfnamefont {C.}~\bibnamefont {Putzke}}, \bibinfo {author} {\bibfnamefont
  {T.}~\bibnamefont {Helm}}, \bibinfo {author} {\bibfnamefont {K.~R.}\
  \bibnamefont {Shirer}}, \bibinfo {author} {\bibfnamefont {Y.-S.}\
  \bibnamefont {Li}}, \bibinfo {author} {\bibfnamefont {K.~A.}\ \bibnamefont
  {Modic}}, \bibinfo {author} {\bibfnamefont {M.}~\bibnamefont {Nicklas}},
  \bibinfo {author} {\bibfnamefont {M.}~\bibnamefont {König}}, \bibinfo
  {author} {\bibfnamefont {D.}~\bibnamefont {Low}}, \bibinfo {author}
  {\bibfnamefont {S.}~\bibnamefont {Ghosh}}, \bibinfo {author} {\bibfnamefont
  {A.~P.}\ \bibnamefont {Mackenzie}}, \bibinfo {author} {\bibfnamefont
  {F.}~\bibnamefont {Arnold}}, \bibinfo {author} {\bibfnamefont
  {E.}~\bibnamefont {Hassinger}}, \bibinfo {author} {\bibfnamefont {R.~D.}\
  \bibnamefont {{McDonald}}}, \bibinfo {author} {\bibfnamefont {L.~E.}\
  \bibnamefont {Winter}}, \bibinfo {author} {\bibfnamefont {E.~D.}\
  \bibnamefont {Bauer}}, \bibinfo {author} {\bibfnamefont {F.}~\bibnamefont
  {Ronning}}, \bibinfo {author} {\bibfnamefont {B.~J.}\ \bibnamefont
  {Ramshaw}}, \bibinfo {author} {\bibfnamefont {K.~C.}\ \bibnamefont
  {Nowack}},\ and\ \bibinfo {author} {\bibfnamefont {P.~J.~W.}\ \bibnamefont
  {Moll}},\ }\bibfield  {title} {\bibinfo {title} {{Spatial control of
  heavy-fermion superconductivity in CeIrIn$_5$}},\ }\href
  {https://doi.org/10.1126/science.aao6640} {\bibfield  {journal} {\bibinfo
  {journal} {Science}\ }\textbf {\bibinfo {volume} {366}},\ \bibinfo {pages}
  {221} (\bibinfo {year} {2019})}\BibitemShut {NoStop}%
\bibitem [{\citenamefont {Liu}\ and\ \citenamefont
  {Maynard}(2012)}]{liu2012measuring}%
  \BibitemOpen
  \bibfield  {author} {\bibinfo {author} {\bibfnamefont {G.}~\bibnamefont
  {Liu}}\ and\ \bibinfo {author} {\bibfnamefont {J.}~\bibnamefont {Maynard}},\
  }\bibfield  {title} {\bibinfo {title} {Measuring elastic constants of
  arbitrarily shaped samples using resonant ultrasound spectroscopy},\ }\href
  {https://doi.org/https://doi.org/10.1121/1.3677259} {\bibfield  {journal}
  {\bibinfo  {journal} {The Journal of the Acoustical Society of America}\
  }\textbf {\bibinfo {volume} {131}},\ \bibinfo {pages} {2068} (\bibinfo {year}
  {2012})}\BibitemShut {NoStop}%
\bibitem [{\citenamefont {Plesek}\ \emph {et~al.}(2004)\citenamefont {Plesek},
  \citenamefont {Kolman},\ and\ \citenamefont {Landa}}]{plesek2004using}%
  \BibitemOpen
  \bibfield  {author} {\bibinfo {author} {\bibfnamefont {J.}~\bibnamefont
  {Plesek}}, \bibinfo {author} {\bibfnamefont {R.}~\bibnamefont {Kolman}},\
  and\ \bibinfo {author} {\bibfnamefont {M.}~\bibnamefont {Landa}},\ }\bibfield
   {title} {\bibinfo {title} {Using finite element method for the determination
  of elastic moduli by resonant ultrasound spectroscopy},\ }\href
  {https://doi.org/https://doi.org/10.1121/1.1760800} {\bibfield  {journal}
  {\bibinfo  {journal} {The Journal of the Acoustical Society of America}\
  }\textbf {\bibinfo {volume} {116}},\ \bibinfo {pages} {282} (\bibinfo {year}
  {2004})}\BibitemShut {NoStop}%
\bibitem [{\citenamefont {Ramshaw}\ \emph {et~al.}(2015)\citenamefont
  {Ramshaw}, \citenamefont {Shekhter}, \citenamefont {McDonald}, \citenamefont
  {Betts}, \citenamefont {Mitchell}, \citenamefont {Tobash}, \citenamefont
  {Mielke}, \citenamefont {Bauer},\ and\ \citenamefont
  {Migliori}}]{Ramshaw2015a}%
  \BibitemOpen
  \bibfield  {author} {\bibinfo {author} {\bibfnamefont {B.~J.}\ \bibnamefont
  {Ramshaw}}, \bibinfo {author} {\bibfnamefont {A.}~\bibnamefont {Shekhter}},
  \bibinfo {author} {\bibfnamefont {R.~D.}\ \bibnamefont {McDonald}}, \bibinfo
  {author} {\bibfnamefont {J.~B.}\ \bibnamefont {Betts}}, \bibinfo {author}
  {\bibfnamefont {J.~N.}\ \bibnamefont {Mitchell}}, \bibinfo {author}
  {\bibfnamefont {P.~H.}\ \bibnamefont {Tobash}}, \bibinfo {author}
  {\bibfnamefont {C.~H.}\ \bibnamefont {Mielke}}, \bibinfo {author}
  {\bibfnamefont {E.~D.}\ \bibnamefont {Bauer}},\ and\ \bibinfo {author}
  {\bibfnamefont {A.}~\bibnamefont {Migliori}},\ }\bibfield  {title} {\bibinfo
  {title} {{Avoided valence transition in a plutonium superconductor}},\ }\href
  {https://doi.org/10.1073/pnas.1421174112} {\bibfield  {journal} {\bibinfo
  {journal} {Proceedings of the National Academy of Sciences of the United
  States of America}\ }\textbf {\bibinfo {volume} {112}},\ \bibinfo {pages}
  {3285} (\bibinfo {year} {2015})}\BibitemShut {NoStop}%
\bibitem [{\citenamefont {Shragai}\ \emph {et~al.}(2023)\citenamefont
  {Shragai}, \citenamefont {Theuss}, \citenamefont {Grissonnanche},\ and\
  \citenamefont {Ramshaw}}]{shragai2023RapidMethodComputing}%
  \BibitemOpen
  \bibfield  {author} {\bibinfo {author} {\bibfnamefont {A.}~\bibnamefont
  {Shragai}}, \bibinfo {author} {\bibfnamefont {F.}~\bibnamefont {Theuss}},
  \bibinfo {author} {\bibfnamefont {G.}~\bibnamefont {Grissonnanche}},\ and\
  \bibinfo {author} {\bibfnamefont {B.~J.}\ \bibnamefont {Ramshaw}},\
  }\bibfield  {title} {\bibinfo {title} {Rapid method for computing the
  mechanical resonances of irregular objects},\ }\href
  {https://doi.org/10.1121/10.0016813} {\bibfield  {journal} {\bibinfo
  {journal} {The Journal of the Acoustical Society of America}\ }\textbf
  {\bibinfo {volume} {153}},\ \bibinfo {pages} {119} (\bibinfo {year}
  {2023})}\BibitemShut {NoStop}%
\bibitem [{\citenamefont {Balakirev}\ \emph {et~al.}(2019)\citenamefont
  {Balakirev}, \citenamefont {Ennaceur}, \citenamefont {Migliori},
  \citenamefont {Maiorov},\ and\ \citenamefont
  {Migliori}}]{balakirev2019resonant}%
  \BibitemOpen
  \bibfield  {author} {\bibinfo {author} {\bibfnamefont {F.~F.}\ \bibnamefont
  {Balakirev}}, \bibinfo {author} {\bibfnamefont {S.~M.}\ \bibnamefont
  {Ennaceur}}, \bibinfo {author} {\bibfnamefont {R.~J.}\ \bibnamefont
  {Migliori}}, \bibinfo {author} {\bibfnamefont {B.}~\bibnamefont {Maiorov}},\
  and\ \bibinfo {author} {\bibfnamefont {A.}~\bibnamefont {Migliori}},\
  }\bibfield  {title} {\bibinfo {title} {Resonant ultrasound spectroscopy: The
  essential toolbox},\ }\href
  {https://doi.org/https://doi.org/10.1063/1.5123165} {\bibfield  {journal}
  {\bibinfo  {journal} {Review of Scientific Instruments}\ }\textbf {\bibinfo
  {volume} {90}},\ \bibinfo {pages} {121401} (\bibinfo {year}
  {2019})}\BibitemShut {NoStop}%
\bibitem [{\citenamefont {Ravindran}\ \emph {et~al.}(1998)\citenamefont
  {Ravindran}, \citenamefont {Fast}, \citenamefont {Korzhavyi}, \citenamefont
  {Johansson}, \citenamefont {Wills},\ and\ \citenamefont
  {Eriksson}}]{doi:10.1063/1.368733}%
  \BibitemOpen
  \bibfield  {author} {\bibinfo {author} {\bibfnamefont {P.}~\bibnamefont
  {Ravindran}}, \bibinfo {author} {\bibfnamefont {L.}~\bibnamefont {Fast}},
  \bibinfo {author} {\bibfnamefont {P.~A.}\ \bibnamefont {Korzhavyi}}, \bibinfo
  {author} {\bibfnamefont {B.}~\bibnamefont {Johansson}}, \bibinfo {author}
  {\bibfnamefont {J.}~\bibnamefont {Wills}},\ and\ \bibinfo {author}
  {\bibfnamefont {O.}~\bibnamefont {Eriksson}},\ }\bibfield  {title} {\bibinfo
  {title} {Density functional theory for calculation of elastic properties of
  orthorhombic crystals: Application to {TiSi$_2$}},\ }\href
  {https://doi.org/10.1063/1.368733} {\bibfield  {journal} {\bibinfo  {journal}
  {Journal of Applied Physics}\ }\textbf {\bibinfo {volume} {84}},\ \bibinfo
  {pages} {4891} (\bibinfo {year} {1998})}\BibitemShut {NoStop}%
\bibitem [{\citenamefont {Weinert}\ \emph {et~al.}(2009)\citenamefont
  {Weinert}, \citenamefont {Schneider}, \citenamefont {Podloucky},\ and\
  \citenamefont {Redinger}}]{FLAPW2009}%
  \BibitemOpen
  \bibfield  {author} {\bibinfo {author} {\bibfnamefont {M.}~\bibnamefont
  {Weinert}}, \bibinfo {author} {\bibfnamefont {G.}~\bibnamefont {Schneider}},
  \bibinfo {author} {\bibfnamefont {R.}~\bibnamefont {Podloucky}},\ and\
  \bibinfo {author} {\bibfnamefont {J.}~\bibnamefont {Redinger}},\ }\bibfield
  {title} {\bibinfo {title} {{FLAPW}: Applications and implementations},\
  }\href {https://doi.org/10.1088/0953-8984/21/8/084201} {\bibfield  {journal}
  {\bibinfo  {journal} {J. Phys. Condens. Matter}\ }\textbf {\bibinfo {volume}
  {21}},\ \bibinfo {pages} {084201} (\bibinfo {year} {2009})}\BibitemShut
  {NoStop}%
\bibitem [{\citenamefont {Perdew}\ \emph {et~al.}(1996)\citenamefont {Perdew},
  \citenamefont {Burke},\ and\ \citenamefont {Ernzerhof}}]{PBE}%
  \BibitemOpen
  \bibfield  {author} {\bibinfo {author} {\bibfnamefont {J.~P.}\ \bibnamefont
  {Perdew}}, \bibinfo {author} {\bibfnamefont {K.}~\bibnamefont {Burke}},\ and\
  \bibinfo {author} {\bibfnamefont {M.}~\bibnamefont {Ernzerhof}},\ }\bibfield
  {title} {\bibinfo {title} {{Generalized Gradient Approximation Made
  Simple}},\ }\href {https://doi.org/10.1103/PhysRevLett.77.3865} {\bibfield
  {journal} {\bibinfo  {journal} {Physical Review Letters}\ }\textbf {\bibinfo
  {volume} {77}},\ \bibinfo {pages} {3865} (\bibinfo {year}
  {1996})}\BibitemShut {NoStop}%
\bibitem [{\citenamefont {Ikeda}\ \emph {et~al.}(2006)\citenamefont {Ikeda},
  \citenamefont {Sakai}, \citenamefont {Aoki}, \citenamefont {Homma},
  \citenamefont {Yamamoto}, \citenamefont {Nakamura}, \citenamefont {Shiokawa},
  \citenamefont {Haga},\ and\ \citenamefont {\={O}nuki}}]{UTe2-lat:2006}%
  \BibitemOpen
  \bibfield  {author} {\bibinfo {author} {\bibfnamefont {S.}~\bibnamefont
  {Ikeda}}, \bibinfo {author} {\bibfnamefont {H.}~\bibnamefont {Sakai}},
  \bibinfo {author} {\bibfnamefont {D.}~\bibnamefont {Aoki}}, \bibinfo {author}
  {\bibfnamefont {Y.}~\bibnamefont {Homma}}, \bibinfo {author} {\bibfnamefont
  {E.}~\bibnamefont {Yamamoto}}, \bibinfo {author} {\bibfnamefont
  {A.}~\bibnamefont {Nakamura}}, \bibinfo {author} {\bibfnamefont
  {Y.}~\bibnamefont {Shiokawa}}, \bibinfo {author} {\bibfnamefont
  {Y.}~\bibnamefont {Haga}},\ and\ \bibinfo {author} {\bibfnamefont
  {Y.}~\bibnamefont {\={O}nuki}},\ }\bibfield  {title} {\bibinfo {title}
  {{Single Crystal Growth and Magnetic Properties of UTe$_2$}},\ }\href
  {https://doi.org/10.1143/JPSJS.75S.116} {\bibfield  {journal} {\bibinfo
  {journal} {Journal of the Physical Society of Japan}\ }\textbf {\bibinfo
  {volume} {75}},\ \bibinfo {pages} {116} (\bibinfo {year} {2006})}\BibitemShut
  {NoStop}%
\bibitem [{\citenamefont {Bell}\ and\ \citenamefont
  {Rupprecht}(1963)}]{bell_elastic_1963}%
  \BibitemOpen
  \bibfield  {author} {\bibinfo {author} {\bibfnamefont {R.~O.}\ \bibnamefont
  {Bell}}\ and\ \bibinfo {author} {\bibfnamefont {G.}~\bibnamefont
  {Rupprecht}},\ }\bibfield  {title} {\bibinfo {title} {{Elastic Constants of
  Strontium Titanate}},\ }\href {https://doi.org/10.1103/PhysRev.129.90}
  {\bibfield  {journal} {\bibinfo  {journal} {Physical Review}\ }\textbf
  {\bibinfo {volume} {129}},\ \bibinfo {pages} {90} (\bibinfo {year}
  {1963})}\BibitemShut {NoStop}%
\bibitem [{\citenamefont {Poindexter}\ and\ \citenamefont
  {Giardini}(1958)}]{poindexter_elastic_1958}%
  \BibitemOpen
  \bibfield  {author} {\bibinfo {author} {\bibfnamefont {E.}~\bibnamefont
  {Poindexter}}\ and\ \bibinfo {author} {\bibfnamefont {A.~A.}\ \bibnamefont
  {Giardini}},\ }\bibfield  {title} {\bibinfo {title} {{Elastic Constants of
  Strontium Titanate (SrTiO$_3$)}},\ }\href
  {https://doi.org/10.1103/PhysRev.110.1069} {\bibfield  {journal} {\bibinfo
  {journal} {Physical Review}\ }\textbf {\bibinfo {volume} {110}},\ \bibinfo
  {pages} {1069} (\bibinfo {year} {1958})}\BibitemShut {NoStop}%
\bibitem [{\citenamefont {Theuss}\ \emph {et~al.}(2022)\citenamefont {Theuss},
  \citenamefont {Ghosh}, \citenamefont {Chen}, \citenamefont {Tchernyshyov},
  \citenamefont {Nakatsuji},\ and\ \citenamefont {Ramshaw}}]{Theuss2022}%
  \BibitemOpen
  \bibfield  {author} {\bibinfo {author} {\bibfnamefont {F.}~\bibnamefont
  {Theuss}}, \bibinfo {author} {\bibfnamefont {S.}~\bibnamefont {Ghosh}},
  \bibinfo {author} {\bibfnamefont {T.}~\bibnamefont {Chen}}, \bibinfo {author}
  {\bibfnamefont {O.}~\bibnamefont {Tchernyshyov}}, \bibinfo {author}
  {\bibfnamefont {S.}~\bibnamefont {Nakatsuji}},\ and\ \bibinfo {author}
  {\bibfnamefont {B.~J.}\ \bibnamefont {Ramshaw}},\ }\bibfield  {title}
  {\bibinfo {title} {{Strong magnetoelastic coupling in ${\mathrm{Mn}}_{3}X$
  ($X=\mathrm{Ge}$, Sn)}},\ }\href
  {https://doi.org/10.1103/PhysRevB.105.174430} {\bibfield  {journal} {\bibinfo
   {journal} {Physical Review B}\ }\textbf {\bibinfo {volume} {105}},\ \bibinfo
  {pages} {174430} (\bibinfo {year} {2022})}\BibitemShut {NoStop}%
\bibitem [{\citenamefont {Storn}\ and\ \citenamefont
  {Price}(1997{\natexlab{a}})}]{storn1997DifferentialEvolutionSimple}%
  \BibitemOpen
  \bibfield  {author} {\bibinfo {author} {\bibfnamefont {R.}~\bibnamefont
  {Storn}}\ and\ \bibinfo {author} {\bibfnamefont {K.}~\bibnamefont {Price}},\
  }\bibfield  {title} {\bibinfo {title} {Differential {{Evolution}}
  \textendash{} {{A Simple}} and {{Efficient Heuristic}} for global
  {{Optimization}} over {{Continuous Spaces}}},\ }\href
  {https://doi.org/10.1023/A:1008202821328} {\bibfield  {journal} {\bibinfo
  {journal} {Journal of Global Optimization}\ }\textbf {\bibinfo {volume}
  {11}},\ \bibinfo {pages} {341} (\bibinfo {year}
  {1997}{\natexlab{a}})}\BibitemShut {NoStop}%
\bibitem [{\citenamefont {Thomas}\ \emph {et~al.}(2020)\citenamefont {Thomas},
  \citenamefont {Santos}, \citenamefont {Christensen}, \citenamefont {Asaba},
  \citenamefont {Ronning}, \citenamefont {Thompson}, \citenamefont {Bauer},
  \citenamefont {Fernandes}, \citenamefont {Fabbris},\ and\ \citenamefont
  {Rosa}}]{thomas_evidence_2020}%
  \BibitemOpen
  \bibfield  {author} {\bibinfo {author} {\bibfnamefont {S.~M.}\ \bibnamefont
  {Thomas}}, \bibinfo {author} {\bibfnamefont {F.~B.}\ \bibnamefont {Santos}},
  \bibinfo {author} {\bibfnamefont {M.~H.}\ \bibnamefont {Christensen}},
  \bibinfo {author} {\bibfnamefont {T.}~\bibnamefont {Asaba}}, \bibinfo
  {author} {\bibfnamefont {F.}~\bibnamefont {Ronning}}, \bibinfo {author}
  {\bibfnamefont {J.~D.}\ \bibnamefont {Thompson}}, \bibinfo {author}
  {\bibfnamefont {E.~D.}\ \bibnamefont {Bauer}}, \bibinfo {author}
  {\bibfnamefont {R.~M.}\ \bibnamefont {Fernandes}}, \bibinfo {author}
  {\bibfnamefont {G.}~\bibnamefont {Fabbris}},\ and\ \bibinfo {author}
  {\bibfnamefont {P.~F.~S.}\ \bibnamefont {Rosa}},\ }\bibfield  {title}
  {\bibinfo {title} {Evidence for a pressure-induced antiferromagnetic quantum
  critical point in intermediate-valence {UTe$_2$}},\ }\href
  {https://doi.org/10.1126/sciadv.abc8709} {\bibfield  {journal} {\bibinfo
  {journal} {Science Advances}\ }\textbf {\bibinfo {volume} {6}},\ \bibinfo
  {pages} {eabc8709} (\bibinfo {year} {2020})}\BibitemShut {NoStop}%
\bibitem [{\citenamefont {Storn}\ and\ \citenamefont
  {Price}(1997{\natexlab{b}})}]{storn1997differential}%
  \BibitemOpen
  \bibfield  {author} {\bibinfo {author} {\bibfnamefont {R.}~\bibnamefont
  {Storn}}\ and\ \bibinfo {author} {\bibfnamefont {K.}~\bibnamefont {Price}},\
  }\bibfield  {title} {\bibinfo {title} {Differential evolution--a simple and
  efficient heuristic for global optimization over continuous spaces},\
  }\href@noop {} {\bibfield  {journal} {\bibinfo  {journal} {Journal of global
  optimization}\ }\textbf {\bibinfo {volume} {11}},\ \bibinfo {pages} {341}
  (\bibinfo {year} {1997}{\natexlab{b}})}\BibitemShut {NoStop}%
\bibitem [{\citenamefont {Virtanen}\ \emph {et~al.}(2020)\citenamefont
  {Virtanen}, \citenamefont {Gommers}, \citenamefont {Oliphant}, \citenamefont
  {Haberland}, \citenamefont {Reddy}, \citenamefont {Cournapeau}, \citenamefont
  {Burovski}, \citenamefont {Peterson}, \citenamefont {Weckesser},
  \citenamefont {Bright}, \citenamefont {{van der Walt}}, \citenamefont
  {Brett}, \citenamefont {Wilson}, \citenamefont {Millman}, \citenamefont
  {Mayorov}, \citenamefont {Nelson}, \citenamefont {Jones}, \citenamefont
  {Kern}, \citenamefont {Larson}, \citenamefont {Carey}, \citenamefont {Polat},
  \citenamefont {Feng}, \citenamefont {Moore}, \citenamefont {{VanderPlas}},
  \citenamefont {Laxalde}, \citenamefont {Perktold}, \citenamefont {Cimrman},
  \citenamefont {Henriksen}, \citenamefont {Quintero}, \citenamefont {Harris},
  \citenamefont {Archibald}, \citenamefont {Ribeiro}, \citenamefont
  {Pedregosa}, \citenamefont {{van Mulbregt}},\ and\ \citenamefont {{SciPy 1.0
  Contributors}}}]{2020SciPy-NMeth}%
  \BibitemOpen
  \bibfield  {author} {\bibinfo {author} {\bibfnamefont {P.}~\bibnamefont
  {Virtanen}}, \bibinfo {author} {\bibfnamefont {R.}~\bibnamefont {Gommers}},
  \bibinfo {author} {\bibfnamefont {T.~E.}\ \bibnamefont {Oliphant}}, \bibinfo
  {author} {\bibfnamefont {M.}~\bibnamefont {Haberland}}, \bibinfo {author}
  {\bibfnamefont {T.}~\bibnamefont {Reddy}}, \bibinfo {author} {\bibfnamefont
  {D.}~\bibnamefont {Cournapeau}}, \bibinfo {author} {\bibfnamefont
  {E.}~\bibnamefont {Burovski}}, \bibinfo {author} {\bibfnamefont
  {P.}~\bibnamefont {Peterson}}, \bibinfo {author} {\bibfnamefont
  {W.}~\bibnamefont {Weckesser}}, \bibinfo {author} {\bibfnamefont
  {J.}~\bibnamefont {Bright}}, \bibinfo {author} {\bibfnamefont {S.~J.}\
  \bibnamefont {{van der Walt}}}, \bibinfo {author} {\bibfnamefont
  {M.}~\bibnamefont {Brett}}, \bibinfo {author} {\bibfnamefont
  {J.}~\bibnamefont {Wilson}}, \bibinfo {author} {\bibfnamefont {K.~J.}\
  \bibnamefont {Millman}}, \bibinfo {author} {\bibfnamefont {N.}~\bibnamefont
  {Mayorov}}, \bibinfo {author} {\bibfnamefont {A.~R.~J.}\ \bibnamefont
  {Nelson}}, \bibinfo {author} {\bibfnamefont {E.}~\bibnamefont {Jones}},
  \bibinfo {author} {\bibfnamefont {R.}~\bibnamefont {Kern}}, \bibinfo {author}
  {\bibfnamefont {E.}~\bibnamefont {Larson}}, \bibinfo {author} {\bibfnamefont
  {C.~J.}\ \bibnamefont {Carey}}, \bibinfo {author} {\bibfnamefont
  {{\.I}.}~\bibnamefont {Polat}}, \bibinfo {author} {\bibfnamefont
  {Y.}~\bibnamefont {Feng}}, \bibinfo {author} {\bibfnamefont {E.~W.}\
  \bibnamefont {Moore}}, \bibinfo {author} {\bibfnamefont {J.}~\bibnamefont
  {{VanderPlas}}}, \bibinfo {author} {\bibfnamefont {D.}~\bibnamefont
  {Laxalde}}, \bibinfo {author} {\bibfnamefont {J.}~\bibnamefont {Perktold}},
  \bibinfo {author} {\bibfnamefont {R.}~\bibnamefont {Cimrman}}, \bibinfo
  {author} {\bibfnamefont {I.}~\bibnamefont {Henriksen}}, \bibinfo {author}
  {\bibfnamefont {E.~A.}\ \bibnamefont {Quintero}}, \bibinfo {author}
  {\bibfnamefont {C.~R.}\ \bibnamefont {Harris}}, \bibinfo {author}
  {\bibfnamefont {A.~M.}\ \bibnamefont {Archibald}}, \bibinfo {author}
  {\bibfnamefont {A.~H.}\ \bibnamefont {Ribeiro}}, \bibinfo {author}
  {\bibfnamefont {F.}~\bibnamefont {Pedregosa}}, \bibinfo {author}
  {\bibfnamefont {P.}~\bibnamefont {{van Mulbregt}}},\ and\ \bibinfo {author}
  {\bibnamefont {{SciPy 1.0 Contributors}}},\ }\bibfield  {title} {\bibinfo
  {title} {{{SciPy} 1.0: Fundamental Algorithms for Scientific Computing in
  Python}},\ }\href {https://doi.org/10.1038/s41592-019-0686-2} {\bibfield
  {journal} {\bibinfo  {journal} {Nature Methods}\ }\textbf {\bibinfo {volume}
  {17}},\ \bibinfo {pages} {261} (\bibinfo {year} {2020})}\BibitemShut
  {NoStop}%
\end{thebibliography}%

%%%%%%%%%%%%% input SI
% \newpage
\clearpage
\onecolumngrid

\section{SUPPLEMENTARY INFORMATION}

\subsection{Bulk Modulus, Young's Moduli, and Poisson's Ratios}
In \autoref{table:bulk modulus} we present the bulk modulus $B$, the Young's moduli $E_{xx}$, $E_{yy}$, $E_{zz}$, and the Poisson's ratios $\nu_{xz}$, $\nu_{yz}$, $\nu_{xy}$ for all measured samples and fits, calculated from the elastic moduli shown in the main text. For cubic \sto, $E_{xx}=E_{yy}=E_{zz}$ and $\nu_{xz}=\nu_{yz}=\nu_{xy}$, whereas for hexagonal \mnge, $E_{xx}=E_{yy}$ and $\nu_{xz}=\nu_{yz}$. For orthorhombic \ute, all moduli differ.

\begin{table*}[hbt!]
	% \centering
	\begin{tabular}{ccccccccc}
		\hline
		\hline
		   &        &           & \multicolumn{3}{c}{Young's Moduli (GPa)}         & \multicolumn{3}{c}{Poisson's Ratios}\\
		   					    \cline{4-9}
	\multicolumn{2}{c}{Sample}       & $B$ (GPa)  & $E_{xx}$ & $E_{yy}$ & $E_{zz}$ & $\nu_{xz}$ & $\nu_{yz}$ & $\nu_{xy}$\\
		\hline
	\multirow{2}{*}{\sto}  & A        & 177       & 271      & -        & -        & 0.24       & -          & -\\
	                       & B        & 174       & 266      & -        & -        & 0.25       & -          & -\\
		\hline
	\multirow{2}{*}{\mnge} & A        & 66        & 115      & -        & 200      & 0.04       & -          & 0.33\\
						   & B        & 65        & 114      & -        & 201      & 0.05       & -          & 0.31\\	
		\hline
    \multirow{2}{*}{\ute}   & 300~K  & 55         & 678      & 126      & 71       & 0.38       & 0.27       & 0.11\\
	    					& 4~K    & 57         & 71       & 131      & 74       & 0.40       & 0.26       & 0.09\\
		\hline
		\hline
	\end{tabular}
	\caption{Bulk modulus and Young's moduli}
	\label{table:bulk modulus}
\end{table*}

\subsection{Density Functional Theory}
In \autoref{table:ute2 DFT} we compare the elastic moduli from density functional theory obtained from calculations during which the atomic coordinates were frozen to experimental values to calculations in which they were optimized. We also compare to calculations presented in the supplement of \cite{girod2022ThermodynamicElectricalTransport}.
\begin{table*}[hbt!]
	% \centering
	\begin{tabular}{cccccccccc}
		\hline
		\hline
	Method & $c_{11}$ & $c_{22}$ & $c_{33}$ & $c_{12}$ & $c_{13}$ & $c_{23}$ & $c_{44}$ & $c_{55}$ & $c_{66}$ \\
		\hline
	coordinates frozen     & 100.2    & 140.0    & 99.3     & 28.7     & 56.4     & 27.1     & 33.8     & 69.4     & 36.1     \\ 
	coordinates optimized  & 95.7     & 136.0    & 89.7     & 28.1     & 46.0     & 26.0     & 28.0     & 57.1     & 31.0     \\ 
	Girod et al. \cite{girod2022ThermodynamicElectricalTransport} & 97.0 & 140.7 & 101.3 & 40.9 & 48.6 & 46.7 & 19.6 & 57.3 & 27.1 \\ 
		\hline
		\hline
	\end{tabular}
	\caption{\textbf{\ute DFT elastic tensor}. Presented are the elastic moduli from density functional theory calculations. The first two rows are our calculations with atomic coordinates frozen to the experimental values and optimized. The last row are values given in the SI of \cite{girod2022ThermodynamicElectricalTransport}, where the atomic coordinates were allowed to relax during minimization.}
	\label{table:ute2 DFT}
\end{table*}

%\clearpage
\subsection{Sample Digitization}
\begin{figure}[hbt!]
	\includegraphics[width=1\columnwidth]{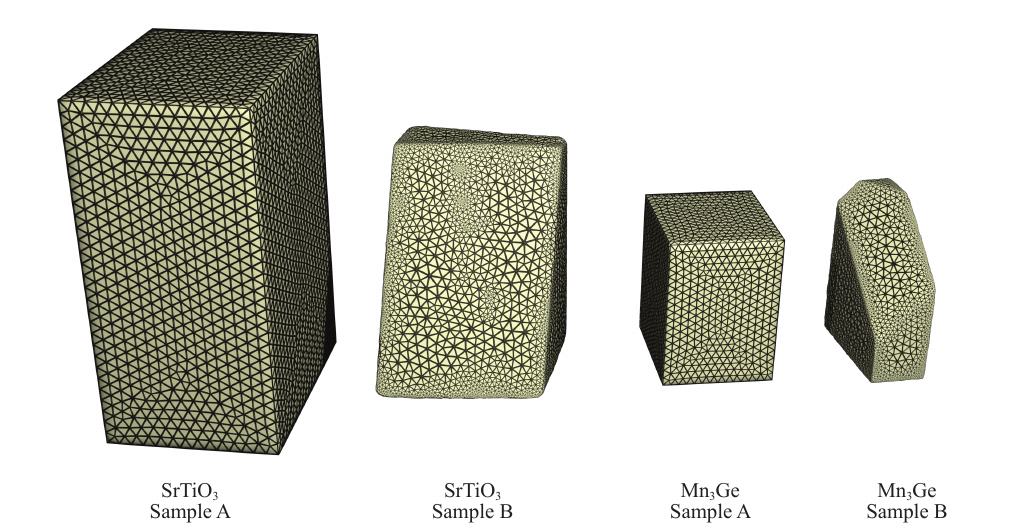}
		\caption{\textbf{CT-Scan Meshes.} Shown are (from left to right) the digitizations for the \sto (samples A, B) and \mnge (samples A, B) samples used for the RUS fits withe the FEM method. Meshes for fits using the SMI method are too dense for individual faces to be identifiable on the shown scale.}
	\label{fig:meshes}
\end{figure}

Three-dimensional digitizations of the samples were obtained with a Zeiss Xradia Versa XRM-520 X-ray nano-CT. The average mesh size varies between samples and RUS solver methods (i.e. SMI vs. FEM). The meshes used for all FEM fits are shown in \autoref{fig:meshes}, and the average mesh size for all fits (FEM and SMI) are given in \autoref{table:mesh size}. We used smaller meshes for irregularly-shaped samples (compared to regularly shaped ones) as well as for fits using the SMI method (compared to the FEM method). The latter is because a smaller mesh size increases the time for a fit to converge significantly for the FEM method, but leaves it almost unaffected for the SMI method \cite{shragai2023RapidMethodComputing}.

% \begin{table*}[hbt!]
% 	% \centering
% 	\begin{tabular}{cccccccccc}
% 	\hline
% 	\hline
% 	Method & \sto     & \sto     & \mnge    & \mnge    & \ute \\
%            & sample A & sample B & sample A & sample B &      \\
%     \hline
%     FEM    &  4600    & 680      & 1700     & 2300     & -    \\
%     SMI    &  1500    & 10       & 240      & 10       & 38   \\
%     \hline
%     \hline
% 	\end{tabular}
% 	\caption{\textbf{Mesh Size}. Average size of a face (in $\mu$m$^2$) in the meshes used in the RUS fits.}
% 	\label{table:mesh size}
% \end{table*}

\begin{table*}[hbt!]
	% \centering
	\begin{tabular}{cccccccccc}
	\hline
	\hline
	Method & \sto     & \sto     & \mnge    & \mnge    & \ute \\
           & sample A & sample B & sample A & sample B &      \\
    \hline
    FEM    &  84      & 56       & 52       & 40       & -    \\
    SMI    &  60      & 5        & 23       & 5        & 9    \\
    \hline
    \hline
	\end{tabular}
	\caption{\textbf{Mesh Size}. Average distance between two vertices (in $\mu$m) in the meshes used in the RUS fits.}
	\label{table:mesh size}
\end{table*}

%\clearpage
\subsection{Differential Evolution Algorithm}

We use a differential evolution global minimizer to extract the elastic moduli from the experimental resonance spectra \cite{storn1997differential, 2020SciPy-NMeth, Ramshaw2015a}. A differential evolution algorithm is initialized with its first generation by randomly generating $N$ sets of elastic moduli within given bounds. We call these sets parent sets, $x_i^{parent}$, $i\in \left[1, 2, ..., N \right]$. Here every $x_i^{parent}$ is a vector containing the independent elastic moduli of a given material. A resonance spectrum is then calculated for all parent sets (see discussion below on details on how resonances are calculated), and compared to the experimental resonance frequencies. The best set of elastic moduli $x^{best}$ is characterized by the lowest $\chi^2$ value. The second generation of the differential evolution algorithm is then created in three steps: First, a mutated set of elastic moduli $x_i^{mut}$ is constructed for every parent set $x_i^{parent}$ through
\begin{equation}
    x_i^{mut} = x^{best} + \epsilon \left( x_j^{parent}+x_k^{parent} \right).
\end{equation}
\par Here, $j,k \in \left[ 1, N \right]$ are random indices, and the mutation parameter $\epsilon \in \left[0,2\right]$ ensures a controlled perturbation of $x^{best}$. Secondly, trial sets $x_i^{trial}$ are created from each mutated parameter set. For this procedure, each element $x_{i,j}^{trial}$---i.e. each elastic modulus---of a given trial set is constructed as
\begin{equation}
	x_{i,j}^{trial} = 
	\begin{cases}
		x_{i,j}^{mut}    & \mathrm{if}~rand(j)\leq p~\mathrm{or}~j=d\\
		x_{i,j}^{parent} & \mathrm{if}~rand(j) > p
	\end{cases},
\end{equation}
where $rand(j)\in \left[0,1\right]$ is a random number, $p \in \left[0,1\right]$ is the crossover probability, and $d$ is the number of elastic moduli.
In the last step, each original parent set $x_i^{parent}$ is compared to its related trial set $x_i^{trial}$ and the set with smallest $\chi^2$ value is chosen for the next generation. This procedure is repeated until the standard deviation of the residuals of a given generation falls below a predetermined tolerance level. The quality of this fit can be improved further by performing a gradient descent fit, using the results of the differential evolution algorithm as starting parameters.

%\clearpage
\subsection{RUS Fit Uncertainties}
The main sources of uncertainties in our fits are due to small deviations from the weak-coupling approximation used in our analysis and from small uncertainties in the alignment between the crystal axes and the sample mesh.

\begin{figure}[hbt!]
	\includegraphics[width=1\columnwidth]{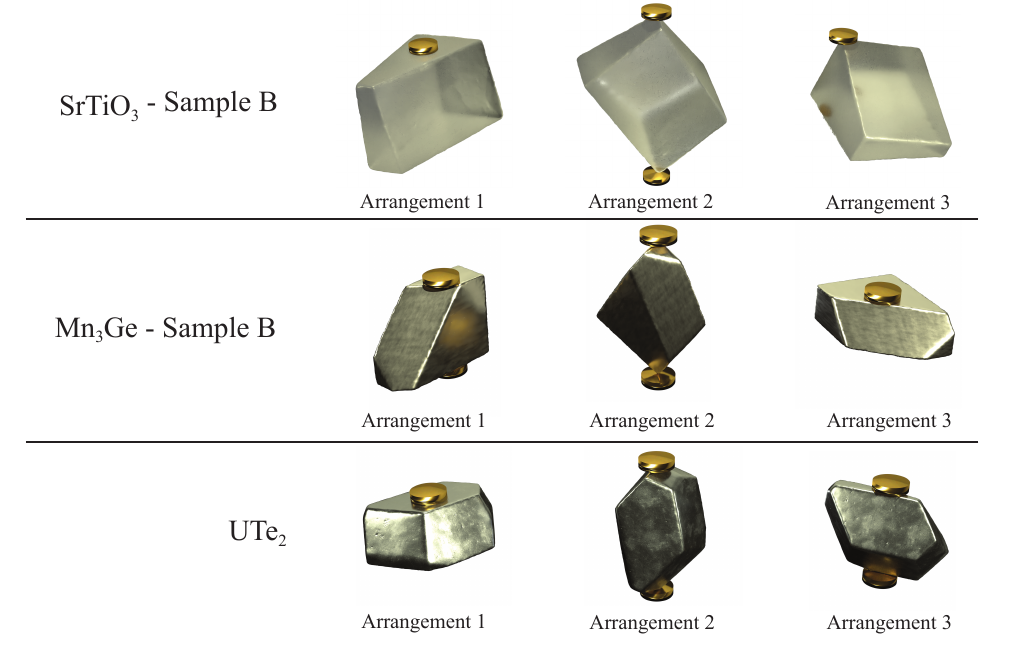}
		\caption{\textbf{RUS Sample Arrangements}. Shown are different arrangements of the samples in the RUS apparatus. The golden discs represent the ultrasonic transducers (the bottom transducer is occluded by the sample in some images). The arrangement number relates to the correct row in \autoref{table:remounting results}.}
	\label{fig:mountings}
\end{figure}

The former is caused by the weight of the cantilever holding the top ultrasonic transducer and causes the measured resonance frequencies to weakly depend on how the sample is mounted in the RUS apparatus. To estimate the resulting uncertainty in elastic moduli, we arranged all irregularly-shaped samples in three different ways between the transducers and re-performed RUS measurements and fits. \autoref{fig:mountings} illustrates the different arrangements of samples in the setup and the rows in \autoref{table:remounting results} show the respective elastic moduli (note that arrangement 2 is how the samples were mounted for the results shown in the main paper). This analysis leads to an average uncertainty of 1.7~GPa, 2.3~GPa, 0.5~GPa for \sto sample B, \mnge sample B, and the \ute sample at 300~K, respectively. The uncertainties are about a factor of 4 larger for the \sto and \mnge samples since they were measured in a RUS apparatus with a cantilever weighing about 5~g, while the \ute spectra were recorded in an apparatus with a cantilever weighing about 0.5~g.

The latter contribution originates in an uncertainty of about 1 degree to which we can align the crystal axes and the sample mesh using Laue backreflection diffractometry. Refitting our RUS resonance spectra with a sample mesh rotated by 1 degree results in average uncertainties of 0.4~GPa, 2~GPa, 0.5~GPa, for the elastic moduli of \sto sample B, \mnge sample B, and \ute at 300~K, respectively. Individual error bars for each elastic modulus due to sample misalignment are given in \autoref{table:remounting results}.

\begin{table*}[hbt!]
	% \centering
	\begin{tabular}{ccccccccccc}
	\hline
	\hline
	Sample                        & Arrangement & $c_{11}$ & $c_{22}$ & $c_{33}$ & $c_{12}$ & $c_{13}$ & $c_{23}$ & $c_{44}$ & $c_{55}$ & $c_{66}$ \\
    \hline
    \multirow{3}{*}{\sto (B)}     & 1           & 313.8         & -        & -        & 100.3         & -        & -        & 122.62          & -        & -        \\
                                  & 2           & 316.7$\pm$0.5 & -        & -        & 102.9$\pm$0.7 & -        & -        & 121.95$\pm$0.05 & -        & -        \\
                                  & 3           & 315.8         & -        & -        & 100.2         & -        & -        & 122.30          & -        & -        \\
    \hline
    \multirow{3}{*}{\mnge (B)}    & 1           & 126       & -        & 200       & 38       & 12       & -        & 48.7         & -        & -        \\
                                  & 2           & 127$\pm$1 & -        & 203$\pm$2 & 40$\pm$2 & 14$\pm$5 & -        & 48.7$\pm$0.5 & -        & -        \\
                                  & 3           & 129       & -        & 196       & 42       & 18       & -        & 49.6         & -        & -        \\
    \hline
    \multirow{3}{*}{\ute (300~K)} & 1           & 84.4          & 139.0         & 92.0         & 25.6         & 38.1         & 32.0         & 26.9         & 52.3         & 29.4     \\
                                  & 2           & 84.7$\pm$0.44 & 139.5$\pm$0.8 & 91.1$\pm$1.1 & 26.8$\pm$1.0 & 38.1$\pm$0.5 & 31.6$\pm$0.4 & 26.9$\pm$0.2 & 52.4$\pm$0.2 & 29.7$\pm$0.2     \\
                                  & 3           & 83.8          & 140.3         & 90.6         & 25.5         & 37.2         & 31.6         & 27.1         & 52.3         & 29.4     \\
    \hline
    \hline
	\end{tabular}
	\caption{\textbf{RUS Uncertainty Analysis}. Fit results for the different arrangements of all irregularly-shaped samples shown in \autoref{fig:mountings} to estimate the uncertainty introduced by the weight of the cantilever holding the top ultrasonic transducer. For each sample and arrangement 2, we also show the errors caused by a 1 degree misalignment between the crystal axes and the sample mesh. All fits were performed with the SMI method.}
	\label{table:remounting results}
\end{table*}

%\clearpage
\subsection{RUS Resonance Spectra and Fit Results}
We show lists of experimental resonance frequencies $f_{exp}$ along with the calculated frequencies $f_{calc}$ corresponding to the RUS fit results in the main text. The lowest three resonances are excluded from all fits.
\begin{longtable}{|c|c|cc|cc|cc|}
	\caption{\sto sample A (regular shape)}\\
	\hline
	      &                 & \multicolumn{2}{c|}{RPR}     & \multicolumn{2}{c|}{SMI}     & \multicolumn{2}{c|}{FEM}\\ 
							\cline{3-8}
	Index & $f_{exp}$ (MHz) & $f_{calc}$ (MHz) & diff (\%) & $f_{calc}$ (MHz) & diff (\%) & $f_{calc}$ (MHz) & diff (\%) \\ 
	\hline
	\endfirsthead
	\multicolumn{8}{r}{Table 2 continued.}\\
	\hline
		&                 & \multicolumn{2}{c|}{RPR}     & \multicolumn{2}{c|}{SMI}     & \multicolumn{2}{c|}{FEM}\\ 
						  \cline{3-8}
	Index & $f_{exp}$ (MHz) & $f_{calc}$ (MHz) & diff (\%) & $f_{calc}$ (MHz) & diff (\%) & $f_{calc}$ (MHz) & diff (\%) \\ 
	\hline
	\endhead
	\hline
	\multicolumn{8}{r}{Table 2 continued on next page.}
	\endfoot
	\hline
	\endlastfoot
	1 & 0.71037 & 0.70491 & -     & 0.70491 & -     & 0.70489 & -     \\
    2 & 0.79197 & 0.79156 & -     & 0.79156 & -     & 0.79155 & -     \\
    3 & 0.89772 & 0.90185 & -     & 0.90185 & -     & 0.90184 & -     \\
    4 & 1.17353 & 1.17781 & 0.363 & 1.17781 & 0.363 & 1.17779 & 0.361 \\
    5 & 1.24570 & 1.24714 & 0.116 & 1.24714 & 0.116 & 1.24711 & 0.113 \\
    6 & 1.33132 & 1.32506 & 0.473 & 1.32506 & 0.473 & 1.32504 & 0.474 \\
    7 & 1.37709 & 1.38138 & 0.310 & 1.38138 & 0.310 & 1.38135 & 0.308 \\
    8 & 1.41735 & 1.40410 & 0.944 & 1.40410 & 0.944 & 1.40408 & 0.945 \\
    9 & 1.50363 & 1.49941 & 0.281 & 1.49941 & 0.281 & 1.49939 & 0.283 \\
    10& 1.55724 & 1.55327 & 0.256 & 1.55327 & 0.256 & 1.55325 & 0.257 \\
    11& 1.59727 & 1.58515 & 0.765 & 1.58515 & 0.765 & 1.58513 & 0.766 \\
    12& 1.60045 & 1.58865 & 0.743 & 1.58865 & 0.743 & 1.58863 & 0.744 \\
    13& 1.71219 & 1.71178 & 0.024 & 1.71178 & 0.024 & 1.71175 & 0.025 \\
    14& 1.77642 & 1.77451 & 0.108 & 1.77451 & 0.108 & 1.77449 & 0.109 \\
    15& 1.78574 & 1.79251 & 0.377 & 1.79251 & 0.377 & 1.79247 & 0.375 \\
    16& 1.83710 & 1.84222 & 0.278 & 1.84222 & 0.277 & 1.84219 & 0.276 \\
    17& 1.85498 & 1.85528 & 0.016 & 1.85528 & 0.016 & 1.85524 & 0.014 \\
    18& 1.91594 & 1.92356 & 0.396 & 1.92356 & 0.396 & 1.92353 & 0.395 \\
    19& 1.97811 & 1.97315 & 0.251 & 1.97315 & 0.251 & 1.97314 & 0.252 \\
    20& 1.98353 & 1.98541 & 0.095 & 1.98541 & 0.095 & 1.98538 & 0.093 \\
    21& 2.03547 & 2.04563 & 0.496 & 2.04563 & 0.496 & 2.04559 & 0.495 \\
    22& 2.07281 & 2.07985 & 0.338 & 2.07985 & 0.338 & 2.07982 & 0.337 \\
    23& 2.16306 & 2.17696 & 0.638 & 2.17696 & 0.639 & 2.17695 & 0.638 \\
    24& 2.17570 & 2.17867 & 0.136 & 2.17867 & 0.136 & 2.17864 & 0.135 \\
    25& 2.20327 & 2.21156 & 0.375 & 2.21156 & 0.375 & 2.21154 & 0.374 \\
    26& 2.20830 & 2.21607 & 0.350 & 2.21607 & 0.350 & 2.21606 & 0.350 \\
    27& 2.22903 & 2.22728 & 0.078 & 2.22728 & 0.078 & 2.22727 & 0.079 \\
    28& 2.23226 & 2.23763 & 0.240 & 2.23763 & 0.240 & 2.23761 & 0.239 \\
    29& 2.23351 & 2.23846 & 0.221 & 2.23846 & 0.221 & 2.23845 & 0.221 \\
    30& 2.28139 & 2.27258 & 0.388 & 2.27258 & 0.388 & 2.27258 & 0.388 \\
    31& 2.32126 & 2.31051 & 0.465 & 2.31051 & 0.465 & 2.31052 & 0.465 \\
    32& 2.33100 & 2.32182 & 0.395 & 2.32182 & 0.395 & 2.32180 & 0.396 \\
    33& 2.34719 & 2.34083 & 0.272 & 2.34083 & 0.272 & 2.34082 & 0.272 \\
    34& 2.44146 & 2.44675 & 0.216 & 2.44675 & 0.216 & 2.44674 & 0.216 \\
    35& 2.46013 & 2.45143 & 0.355 & 2.45142 & 0.355 & 2.45144 & 0.355 \\
    36& 2.50210 & 2.50958 & 0.298 & 2.50958 & 0.298 & 2.50956 & 0.297 \\
    37& 2.50662 & 2.50968 & 0.122 & 2.50968 & 0.122 & 2.50965 & 0.121 \\
    38& 2.53756 & 2.53887 & 0.051 & 2.53887 & 0.051 & 2.53889 & 0.052 \\
    39& 2.57650 & 2.57679 & 0.011 & 2.57679 & 0.011 & 2.57680 & 0.012 \\
    40& 2.58500 & 2.59496 & 0.384 & 2.59496 & 0.384 & 2.59496 & 0.384 \\
    41& 2.59850 & 2.59922 & 0.028 & 2.59922 & 0.028 & 2.59920 & 0.027 \\
    42& 2.62865 & 2.63115 & 0.095 & 2.63116 & 0.095 & 2.63117 & 0.096 \\
    43& 2.65167 & 2.65229 & 0.023 & 2.65229 & 0.023 & 2.65226 & 0.022 \\
    44& 2.68309 & 2.68338 & 0.011 & 2.68338 & 0.011 & 2.68338 & 0.011 \\
    45& 2.70518 & 2.70901 & 0.141 & 2.70901 & 0.141 & 2.70898 & 0.140 \\
    46& 2.74782 & 2.74881 & 0.036 & 2.74881 & 0.036 & 2.74883 & 0.037 \\
    47& 2.75176 & 2.75107 & 0.025 & 2.75107 & 0.025 & 2.75107 & 0.025 \\
    48& 2.75392 & 2.76091 & 0.253 & 2.76091 & 0.253 & 2.76092 & 0.254 \\
    49& 2.78789 & 2.79403 & 0.220 & 2.79403 & 0.220 & 2.79408 & 0.221 \\
    50& 2.80071 & 2.79544 & 0.189 & 2.79545 & 0.188 & 2.79545 & 0.188 \\
    51& 2.82197 & 2.82346 & 0.053 & 2.82346 & 0.053 & 2.82352 & 0.055 \\
    52& 2.84439 & 2.85166 & 0.255 & 2.85167 & 0.255 & 2.85166 & 0.255 \\
    53& 2.91635 & 2.91928 & 0.101 & 2.91928 & 0.101 & 2.91931 & 0.102 \\
    54& 2.92322 & 2.92473 & 0.051 & 2.92473 & 0.051 & 2.92478 & 0.053 \\
    55& 2.94885 & 2.92904 & 0.677 & 2.92904 & 0.677 & 2.92902 & 0.677 \\
    56& 3.00333 & 2.99929 & 0.135 & 2.99929 & 0.135 & 2.99940 & 0.131 \\
    57& 3.01642 & 3.01070 & 0.190 & 3.01070 & 0.190 & 3.01074 & 0.189 \\
    58& 3.03353 & 3.03246 & 0.035 & 3.03246 & 0.036 & 3.03254 & 0.033 \\
    59& 3.03869 & 3.04328 & 0.151 & 3.04328 & 0.151 & 3.04332 & 0.152 \\
    60& 3.06889 & 3.06572 & 0.103 & 3.06572 & 0.103 & 3.06574 & 0.103 \\
    61& 3.13815 & 3.14283 & 0.149 & 3.14283 & 0.149 & 3.14288 & 0.151 \\
    62& 3.15099 & 3.15332 & 0.074 & 3.15332 & 0.074 & 3.15339 & 0.076 \\
    63& 3.15918 & 3.16111 & 0.061 & 3.16111 & 0.061 & 3.16112 & 0.062 \\
    64& 3.16664 & 3.16975 & 0.098 & 3.16975 & 0.098 & 3.16985 & 0.101 \\
    65& 3.22486 & 3.22895 & 0.127 & 3.22895 & 0.127 & 3.22893 & 0.126 \\
    66& 3.22600 & 3.23110 & 0.158 & 3.23110 & 0.158 & 3.23107 & 0.157 \\
    67& 3.25102 & 3.24020 & 0.334 & 3.24020 & 0.334 & 3.24033 & 0.330 \\
    68& 3.25333 & 3.24905 & 0.132 & 3.24905 & 0.132 & 3.24915 & 0.129 \\
    69& 3.26656 & 3.25129 & 0.470 & 3.25129 & 0.470 & 3.25141 & 0.466 \\
    70& 3.27348 & 3.28127 & 0.237 & 3.28127 & 0.237 & 3.28131 & 0.239 \\
\end{longtable}	 

\begin{longtable}{|c|c|cc|cc|}
	\caption{\sto sample B (irregular shape)}\\
	\hline
	      &                 & \multicolumn{2}{c|}{SMI}     & \multicolumn{2}{c|}{FEM}\\ 
							\cline{3-6}
	Index & $f_{exp}$ (MHz) & $f_{calc}$ (MHz) & diff (\%) & $f_{calc}$ (MHz) & diff (\%) \\ 
	\hline
	\endfirsthead
	\multicolumn{6}{r}{Table 3 continued.}\\
	\hline
		&                 & \multicolumn{2}{c|}{SMI}     & \multicolumn{2}{c|}{FEM}\\ 
						  \cline{3-6}
	Index & $f_{exp}$ (MHz) & $f_{calc}$ (MHz) & diff (\%) & $f_{calc}$ (MHz) & diff (\%) \\ 
	\hline
	\endhead
	\hline
	\multicolumn{6}{r}{Table 3 continued on next page.}
	\endfoot
	\hline
	\endlastfoot
	1 & 1.10869 & 1.09168 & -     & 1.09156 & -    \\
    2 & 1.20480 & 1.19121 & -     & 1.19112 & -    \\
    3 & 1.52331 & 1.50347 & -     & 1.50341 & -    \\
    4 & 1.88059 & 1.86636 & 0.762 & 1.86620 & 0.771\\
    5 & 1.97243 & 1.97144 & 0.050 & 1.97135 & 0.055\\
    6 & 1.98450 & 1.98777 & 0.164 & 1.98756 & 0.154\\
    7 & 2.07997 & 2.07118 & 0.424 & 2.07100 & 0.433\\
    8 & 2.15054 & 2.14467 & 0.274 & 2.14453 & 0.280\\
    9 & 2.23955 & 2.23207 & 0.335 & 2.23186 & 0.344\\
    10& 2.37823 & 2.36840 & 0.415 & 2.36833 & 0.418\\
    11& 2.43553 & 2.44027 & 0.194 & 2.44001 & 0.184\\
    12& 2.58487 & 2.57275 & 0.471 & 2.57258 & 0.478\\
    13& 2.72511 & 2.71927 & 0.215 & 2.71912 & 0.220\\
    14& 2.80098 & 2.79093 & 0.360 & 2.79078 & 0.366\\
    15& 2.83415 & 2.83210 & 0.072 & 2.83199 & 0.077\\
    16& 2.87280 & 2.86311 & 0.339 & 2.86304 & 0.341\\
    17& 2.92701 & 2.91494 & 0.414 & 2.91473 & 0.421\\
    18& 3.07796 & 3.07446 & 0.114 & 3.07432 & 0.118\\
    19& 3.13235 & 3.14863 & 0.517 & 3.14864 & 0.517\\
    20& 3.16734 & 3.16494 & 0.076 & 3.16486 & 0.079\\
    21& 3.18282 & 3.18195 & 0.027 & 3.18191 & 0.029\\
    22& 3.25382 & 3.24307 & 0.331 & 3.24298 & 0.334\\
    23& 3.29615 & 3.29027 & 0.179 & 3.29022 & 0.180\\
    24& 3.31731 & 3.33535 & 0.541 & 3.33524 & 0.538\\
    25& 3.48056 & 3.50474 & 0.690 & 3.50469 & 0.688\\
    26& 3.51006 & 3.52415 & 0.400 & 3.52412 & 0.399\\
    27& 3.65645 & 3.66182 & 0.147 & 3.66178 & 0.145\\
    28& 3.78318 & 3.79193 & 0.231 & 3.79191 & 0.230\\
    29& 3.84523 & 3.84207 & 0.082 & 3.84198 & 0.085\\
    30& 3.89894 & 3.90375 & 0.123 & 3.90368 & 0.122\\
    31& 3.98772 & 3.99729 & 0.239 & 3.99727 & 0.239\\
    32& 4.00064 & 4.00409 & 0.086 & 4.00407 & 0.086\\
    33& 4.06006 & 4.04985 & 0.252 & 4.04979 & 0.254\\
    34& 4.10005 & 4.10447 & 0.108 & 4.10444 & 0.107\\
    35& 4.15255 & 4.14028 & 0.296 & 4.14022 & 0.298\\
    36& 4.18979 & 4.20182 & 0.286 & 4.20174 & 0.284\\
    37& 4.23696 & 4.22814 & 0.209 & 4.22819 & 0.207\\
    38& 4.27449 & 4.28863 & 0.330 & 4.28857 & 0.328\\
    39& 4.30861 & 4.31115 & 0.059 & 4.31124 & 0.061\\
    40& 4.34986 & 4.35327 & 0.078 & 4.35335 & 0.080\\
    41& 4.39819 & 4.39097 & 0.165 & 4.39113 & 0.161\\
    42& 4.46403 & 4.46713 & 0.069 & 4.46719 & 0.071\\
    43& 4.48597 & 4.47481 & 0.249 & 4.47493 & 0.247\\
    44& 4.63845 & 4.62951 & 0.193 & 4.62957 & 0.192\\
    45& 4.68243 & 4.69013 & 0.164 & 4.69022 & 0.166\\
    46& 4.74359 & 4.74395 & 0.008 & 4.74417 & 0.012\\
    47& 4.75697 & 4.76987 & 0.270 & 4.76996 & 0.272\\
    48& 4.80223 & 4.79540 & 0.142 & 4.79552 & 0.140\\
    49& 4.82394 & 4.84266 & 0.387 & 4.84274 & 0.388\\
    50& 4.89389 & 4.90448 & 0.216 & 4.90445 & 0.215\\
    51& 4.93746 & 4.93653 & 0.019 & 4.93677 & 0.014\\
    52& 4.98372 & 4.98686 & 0.063 & 4.98700 & 0.066\\
    53& 5.00140 & 5.00252 & 0.022 & 5.00247 & 0.021\\
    54& 5.03260 & 5.03444 & 0.037 & 5.03457 & 0.039\\
    55& 5.06065 & 5.06660 & 0.117 & 5.06673 & 0.120\\
    56& 5.10892 & 5.10434 & 0.090 & 5.10460 & 0.085\\
    57& 5.12704 & 5.12970 & 0.052 & 5.12991 & 0.056\\
    58& 5.16940 & 5.18029 & 0.210 & 5.18061 & 0.216\\
    59& 5.20187 & 5.23093 & 0.556 & 5.23130 & 0.562\\
    60& 5.25259 & 5.25272 & 0.002 & 5.25278 & 0.004\\
    61& 5.27776 & 5.27884 & 0.020 & 5.27913 & 0.026\\
    62& 5.28863 & 5.29927 & 0.201 & 5.29969 & 0.209\\
    63& 5.33266 & 5.33158 & 0.020 & 5.33170 & 0.018\\
    64& 5.37012 & 5.35725 & 0.240 & 5.35743 & 0.237\\
    65& 5.40014 & 5.39589 & 0.079 & 5.39627 & 0.072\\
    66& 5.43224 & 5.43571 & 0.064 & 5.43599 & 0.069\\
    67& 5.45845 & 5.45699 & 0.027 & 5.45728 & 0.021\\
    68& 5.49232 & 5.49778 & 0.099 & 5.49816 & 0.106\\
    69& 5.54920 & 5.55358 & 0.079 & 5.55405 & 0.087\\
    70& 5.58061 & 5.58871 & 0.145 & 5.58914 & 0.153\\
\end{longtable}

\begin{longtable}{|c|c|cc|cc|cc|}
	\caption{\mnge sample A (regular shape)}\\
	\hline
	      &                 & \multicolumn{2}{c|}{RPR}     & \multicolumn{2}{c|}{SMI}     & \multicolumn{2}{c|}{FEM}\\ 
							\cline{3-8}
	Index & $f_{exp}$ (MHz) & $f_{calc}$ (MHz) & diff (\%) & $f_{calc}$ (MHz) & diff (\%) & $f_{calc}$ (MHz) & diff (\%) \\ 
	\hline
	\endfirsthead
	\multicolumn{8}{r}{Table 4 continued.}\\
	\hline
		&                 & \multicolumn{2}{c|}{RPR}     & \multicolumn{2}{c|}{SMI}     & \multicolumn{2}{c|}{FEM}\\ 
						  \cline{3-8}
	Index & $f_{exp}$ (MHz) & $f_{calc}$ (MHz) & diff (\%) & $f_{calc}$ (MHz) & diff (\%) & $f_{calc}$ (MHz) & diff (\%) \\ 
	\hline
	\endhead
	\hline
	\multicolumn{8}{r}{Table 4 continued on next page.}
	\endfoot
	\hline
	\endlastfoot
	1 & 0.96469 & 0.95542 & -     & 0.95544 & -     & 0.95531 & -     \\
    2 & 1.31011 & 1.31762 & -     & 1.31764 & -     & 1.31751 & -     \\
    3 & 1.46748 & 1.46874 & -     & 1.46876 & -     & 1.46855 & -     \\
    4 & 1.50955 & 1.50952 & 0.002 & 1.50951 & 0.002 & 1.50947 & 0.005 \\
    5 & 1.56402 & 1.56137 & 0.169 & 1.56136 & 0.170 & 1.56127 & 0.176 \\
    6 & 1.57830 & 1.57952 & 0.077 & 1.57951 & 0.077 & 1.57945 & 0.073 \\
    7 & 1.58425 & 1.60049 & 1.015 & 1.60051 & 1.016 & 1.60029 & 1.002 \\
    8 & 1.60141 & 1.61340 & 0.743 & 1.61340 & 0.743 & 1.61327 & 0.735 \\
    9 & 1.60582 & 1.61949 & 0.845 & 1.61951 & 0.846 & 1.61938 & 0.838 \\
    10& 1.78631 & 1.80175 & 0.857 & 1.80177 & 0.858 & 1.80162 & 0.850 \\
    11& 1.82036 & 1.81183 & 0.471 & 1.81182 & 0.471 & 1.81169 & 0.478 \\
    12& 1.87303 & 1.87322 & 0.010 & 1.87322 & 0.010 & 1.87313 & 0.005 \\
    13& 1.94132 & 1.93235 & 0.464 & 1.93233 & 0.465 & 1.93222 & 0.471 \\
    14& 1.95112 & 1.95456 & 0.176 & 1.95455 & 0.175 & 1.95451 & 0.173 \\
    15& 2.03329 & 2.02996 & 0.164 & 2.02997 & 0.164 & 2.02986 & 0.169 \\
    16& 2.09609 & 2.08355 & 0.602 & 2.08353 & 0.603 & 2.08339 & 0.610 \\
    17& 2.10555 & 2.10872 & 0.150 & 2.10867 & 0.148 & 2.10854 & 0.142 \\
    18& 2.14587 & 2.12376 & 1.041 & 2.12373 & 1.043 & 2.12363 & 1.047 \\
    19& 2.16239 & 2.17045 & 0.371 & 2.17041 & 0.369 & 2.17050 & 0.374 \\
    20& 2.16541 & 2.17570 & 0.473 & 2.17572 & 0.474 & 2.17548 & 0.463 \\
    21& 2.25310 & 2.26434 & 0.496 & 2.26437 & 0.497 & 2.26423 & 0.491 \\
    22& 2.32002 & 2.30809 & 0.517 & 2.30807 & 0.518 & 2.30797 & 0.522 \\
    23& 2.37049 & 2.35061 & 0.846 & 2.35060 & 0.846 & 2.35052 & 0.849 \\
    24& 2.38275 & 2.38340 & 0.027 & 2.38330 & 0.023 & 2.38334 & 0.024 \\
    25& 2.43586 & 2.41898 & 0.698 & 2.41896 & 0.699 & 2.41890 & 0.701 \\
    26& 2.46256 & 2.47420 & 0.471 & 2.47422 & 0.471 & 2.47413 & 0.468 \\
    27& 2.47301 & 2.47498 & 0.080 & 2.47498 & 0.080 & 2.47483 & 0.073 \\
    28& 2.52384 & 2.55083 & 1.058 & 2.55069 & 1.053 & 2.55063 & 1.051 \\
    29& 2.55464 & 2.57203 & 0.676 & 2.57205 & 0.677 & 2.57190 & 0.671 \\
    30& 2.57758 & 2.59117 & 0.524 & 2.59118 & 0.525 & 2.59105 & 0.520 \\
    31& 2.61331 & 2.62284 & 0.363 & 2.62282 & 0.363 & 2.62273 & 0.359 \\
    32& 2.65457 & 2.64939 & 0.196 & 2.64940 & 0.195 & 2.64940 & 0.195 \\
    33& 2.71919 & 2.72259 & 0.125 & 2.72258 & 0.124 & 2.72248 & 0.121 \\
    34& 2.75160 & 2.74615 & 0.198 & 2.74616 & 0.198 & 2.74602 & 0.203 \\
    35& 2.78098 & 2.77733 & 0.131 & 2.77734 & 0.131 & 2.77722 & 0.135 \\
    36& 2.84224 & 2.83905 & 0.112 & 2.83904 & 0.113 & 2.83904 & 0.113 \\
    37& 2.86041 & 2.86306 & 0.092 & 2.86298 & 0.090 & 2.86289 & 0.086 \\
    38& 2.87166 & 2.86844 & 0.112 & 2.86847 & 0.111 & 2.86843 & 0.113 \\
    39& 2.91103 & 2.92610 & 0.515 & 2.92611 & 0.516 & 2.92617 & 0.518 \\
    40& 2.93140 & 2.92769 & 0.127 & 2.92770 & 0.126 & 2.92756 & 0.131 \\
    41& 2.93455 & 2.93191 & 0.090 & 2.93190 & 0.090 & 2.93189 & 0.091 \\
    42& 2.94156 & 2.94237 & 0.027 & 2.94237 & 0.028 & 2.94228 & 0.024 \\
    43& 3.04438 & 3.05124 & 0.225 & 3.05123 & 0.224 & 3.05109 & 0.220 \\
    44& 3.10644 & 3.09616 & 0.332 & 3.09616 & 0.332 & 3.09622 & 0.330 \\
    45& 3.12912 & 3.13303 & 0.125 & 3.13303 & 0.125 & 3.13297 & 0.123 \\
    46& 3.19343 & 3.17795 & 0.487 & 3.17796 & 0.487 & 3.17801 & 0.485 \\
    47& 3.21141 & 3.21040 & 0.031 & 3.21041 & 0.031 & 3.21032 & 0.034 \\
    48& 3.28658 & 3.28210 & 0.136 & 3.28212 & 0.136 & 3.28219 & 0.134 \\
    49& 3.30955 & 3.29828 & 0.342 & 3.29828 & 0.342 & 3.29851 & 0.335 \\
    50& 3.31565 & 3.32261 & 0.209 & 3.32261 & 0.209 & 3.32258 & 0.209 \\
    51& 3.35754 & 3.37386 & 0.484 & 3.37387 & 0.484 & 3.37382 & 0.482 \\
    52& 3.38287 & 3.37789 & 0.147 & 3.37786 & 0.148 & 3.37795 & 0.146 \\
    53& 3.44121 & 3.45795 & 0.484 & 3.45795 & 0.484 & 3.45796 & 0.484 \\
    54& 3.47190 & 3.46067 & 0.324 & 3.46069 & 0.324 & 3.46107 & 0.313 \\
    55& 3.49421 & 3.48695 & 0.208 & 3.48694 & 0.208 & 3.48696 & 0.208 \\
    56& 3.49812 & 3.50059 & 0.071 & 3.50057 & 0.070 & 3.50061 & 0.071 \\
    57& 3.49964 & 3.51232 & 0.361 & 3.51236 & 0.362 & 3.51229 & 0.360 \\
    58& 3.54746 & 3.53238 & 0.427 & 3.53237 & 0.427 & 3.53259 & 0.421 \\
    59& 3.57278 & 3.56487 & 0.222 & 3.56488 & 0.222 & 3.56526 & 0.211 \\
    60& 3.58496 & 3.57191 & 0.365 & 3.57191 & 0.365 & 3.57200 & 0.363 \\
    61& 3.58767 & 3.59720 & 0.265 & 3.59714 & 0.263 & 3.59709 & 0.262 \\
    62& 3.60231 & 3.60308 & 0.022 & 3.60311 & 0.022 & 3.60304 & 0.020 \\
    63& 3.60760 & 3.60521 & 0.066 & 3.60521 & 0.066 & 3.60540 & 0.061 \\
    64& 3.61308 & 3.60655 & 0.181 & 3.60655 & 0.181 & 3.60663 & 0.179 \\
    65& 3.61436 & 3.61431 & 0.001 & 3.61432 & 0.001 & 3.61455 & 0.005 \\
    66& 3.61865 & 3.62068 & 0.056 & 3.62069 & 0.057 & 3.62084 & 0.060 \\
    67& 3.64765 & 3.64158 & 0.167 & 3.64152 & 0.168 & 3.64153 & 0.168 \\
    68& 3.67018 & 3.64580 & 0.669 & 3.64580 & 0.669 & 3.64602 & 0.663 \\
    69& 3.71718 & 3.70179 & 0.416 & 3.70179 & 0.416 & 3.70203 & 0.409 \\
    70& 3.72327 & 3.72386 & 0.016 & 3.72386 & 0.016 & 3.72412 & 0.023 \\
    71& 3.78761 & 3.79633 & 0.230 & 3.79635 & 0.230 & 3.79659 & 0.237 \\
    72& 3.81140 & 3.80554 & 0.154 & 3.80556 & 0.153 & 3.80579 & 0.147 \\
    73& 3.84430 & 3.82293 & 0.559 & 3.82295 & 0.559 & 3.82310 & 0.554 \\
    74& 3.86988 & 3.84886 & 0.546 & 3.84885 & 0.546 & 3.84915 & 0.539 \\
    75& 3.91723 & 3.91340 & 0.098 & 3.91339 & 0.098 & 3.91352 & 0.095 \\
    76& 3.94563 & 3.92345 & 0.565 & 3.92345 & 0.565 & 3.92395 & 0.552 \\
    77& 3.97908 & 3.97784 & 0.031 & 3.97787 & 0.030 & 3.97802 & 0.027 \\
    78& 3.99942 & 3.99986 & 0.011 & 3.99985 & 0.011 & 4.00031 & 0.022 \\
    79& 4.01915 & 4.00900 & 0.253 & 4.00903 & 0.252 & 4.00943 & 0.242 \\
    80& 4.03495 & 4.02355 & 0.284 & 4.02355 & 0.283 & 4.02443 & 0.262 \\
    81& 4.04595 & 4.06045 & 0.357 & 4.06046 & 0.357 & 4.06074 & 0.364 \\
    82& 4.05758 & 4.06567 & 0.199 & 4.06558 & 0.197 & 4.06568 & 0.199 \\
    83& 4.06491 & 4.08637 & 0.525 & 4.08631 & 0.524 & 4.08631 & 0.524 \\
    84& 4.0780  & 4.09008 & 0.295 & 4.09004 & 0.294 & 4.09015 & 0.297 \\
\end{longtable}

\begin{longtable}{|c|c|cc|cc|}
	\caption{\mnge sample B (irregular shape)}\\
	\hline
	      &                 & \multicolumn{2}{c|}{SMI}     & \multicolumn{2}{c|}{FEM}\\ 
							\cline{3-6}
	Index & $f_{exp}$ (MHz) & $f_{calc}$ (MHz) & diff (\%) & $f_{calc}$ (MHz) & diff (\%) \\ 
	\hline
	\endfirsthead
	\multicolumn{6}{r}{Table 5 continued.}\\
	\hline
		&                 & \multicolumn{2}{c|}{SMI}     & \multicolumn{2}{c|}{FEM}\\ 
						  \cline{3-6}
	Index & $f_{exp}$ (MHz) & $f_{calc}$ (MHz) & diff (\%) & $f_{calc}$ (MHz) & diff (\%) \\ 
	\hline
	\endhead
	\hline
	\multicolumn{6}{r}{Table 5 continued on next page.}
	\endfoot
	\hline
	\endlastfoot
	1 & 0.77100 & 0.72927 & -     & 0.72932 & -     \\
    2 & 1.08478 & 1.06421 & -     & 1.06356 & -     \\
    3 & 1.22382 & 1.21029 & -     & 1.21031 & -     \\
    4 & 1.33437 & 1.32618 & 0.618 & 1.32623 & 0.614 \\
    5 & 1.44953 & 1.43028 & 1.346 & 1.43030 & 1.344 \\
    6 & 1.60718 & 1.59563 & 0.723 & 1.59565 & 0.722 \\
    7 & 1.86078 & 1.84572 & 0.816 & 1.84544 & 0.831 \\
    8 & 1.88524 & 1.88848 & 0.172 & 1.88804 & 0.148 \\
    9 & 2.02059 & 2.01970 & 0.044 & 2.01929 & 0.064 \\
    10& 2.06617 & 2.03724 & 1.420 & 2.03660 & 1.452 \\
    11& 2.12700 & 2.11597 & 0.521 & 2.11593 & 0.523 \\
    12& 2.30712 & 2.29890 & 0.358 & 2.29885 & 0.360 \\
    13& 2.33951 & 2.33223 & 0.312 & 2.33215 & 0.316 \\
    14& 2.44134 & 2.43977 & 0.064 & 2.43948 & 0.076 \\
    15& 2.48376 & 2.48460 & 0.034 & 2.48477 & 0.040 \\
    16& 2.66501 & 2.66902 & 0.150 & 2.66862 & 0.135 \\
    17& 2.75780 & 2.74510 & 0.463 & 2.74453 & 0.484 \\
    18& 2.78254 & 2.78677 & 0.152 & 2.78699 & 0.160 \\
    19& 2.83647 & 2.82217 & 0.507 & 2.82196 & 0.514 \\
    20& 2.97943 & 2.99342 & 0.467 & 2.99356 & 0.472 \\
    21& 3.01159 & 3.01770 & 0.203 & 3.01823 & 0.220 \\
    22& 3.07402 & 3.09174 & 0.573 & 3.09132 & 0.560 \\
    23& 3.13148 & 3.13824 & 0.215 & 3.13768 & 0.198 \\
    24& 3.16311 & 3.17632 & 0.416 & 3.17606 & 0.408 \\
    25& 3.24362 & 3.25436 & 0.330 & 3.25461 & 0.338 \\
    26& 3.25447 & 3.27058 & 0.492 & 3.27034 & 0.485 \\
    27& 3.30929 & 3.30305 & 0.189 & 3.30309 & 0.188 \\
    28& 3.40908 & 3.42759 & 0.540 & 3.42754 & 0.539 \\
    29& 3.46136 & 3.46796 & 0.190 & 3.46799 & 0.191 \\
    30& 3.53828 & 3.52248 & 0.449 & 3.52213 & 0.459 \\
    31& 3.58063 & 3.59778 & 0.477 & 3.59745 & 0.468 \\
    32& 3.64885 & 3.64490 & 0.108 & 3.64495 & 0.107 \\
    33& 3.70451 & 3.68606 & 0.500 & 3.68572 & 0.510 \\
    34& 3.73124 & 3.74476 & 0.361 & 3.74458 & 0.356 \\
    35& 3.77514 & 3.78817 & 0.344 & 3.78828 & 0.347 \\
    36& 3.81664 & 3.81972 & 0.081 & 3.82010 & 0.091 \\
    37& 3.84659 & 3.86236 & 0.408 & 3.86303 & 0.426 \\
    38& 3.89301 & 3.90315 & 0.260 & 3.90408 & 0.283 \\
    39& 3.92147 & 3.92013 & 0.034 & 3.92011 & 0.035 \\
    40& 4.00155 & 4.01985 & 0.455 & 4.01957 & 0.448 \\
    41& 4.05232 & 4.04810 & 0.104 & 4.04792 & 0.109 \\
    42& 4.07279 & 4.07673 & 0.096 & 4.07692 & 0.101 \\
    43& 4.11185 & 4.10555 & 0.154 & 4.10515 & 0.163 \\
    44& 4.16568 & 4.12779 & 0.918 & 4.12750 & 0.925 \\
    45& 4.17169 & 4.15185 & 0.478 & 4.15181 & 0.479 \\
    46& 4.23171 & 4.23428 & 0.061 & 4.23422 & 0.059 \\
    47& 4.25632 & 4.26175 & 0.127 & 4.26195 & 0.132 \\
    48& 4.28696 & 4.29318 & 0.145 & 4.29333 & 0.148 \\
    49& 4.33501 & 4.33233 & 0.062 & 4.33252 & 0.057 \\
    50& 4.39232 & 4.39776 & 0.124 & 4.39806 & 0.131 \\
    51& 4.44617 & 4.44492 & 0.028 & 4.44480 & 0.031 \\
    52& 4.45932 & 4.45990 & 0.013 & 4.45953 & 0.005 \\
    53& 4.47269 & 4.48323 & 0.235 & 4.48383 & 0.248 \\
    54& 4.50973 & 4.51226 & 0.056 & 4.51274 & 0.067 \\
    55& 4.54993 & 4.53911 & 0.239 & 4.53896 & 0.242 \\
    56& 4.57564 & 4.59124 & 0.340 & 4.59160 & 0.348 \\
    57& 4.61763 & 4.59924 & 0.400 & 4.59942 & 0.396 \\
    58& 4.64177 & 4.63850 & 0.070 & 4.63892 & 0.061 \\
    59& 4.67340 & 4.67170 & 0.036 & 4.67168 & 0.037 \\
    60& 4.67725 & 4.71140 & 0.725 & 4.71268 & 0.752 \\
    61& 4.74328 & 4.72813 & 0.320 & 4.72823 & 0.318 \\
    62& 4.75405 & 4.76937 & 0.321 & 4.76985 & 0.331 \\
    63& 4.78548 & 4.79885 & 0.279 & 4.79964 & 0.295 \\
    64& 4.83816 & 4.84783 & 0.199 & 4.84855 & 0.214 \\
    65& 4.87454 & 4.89056 & 0.328 & 4.89094 & 0.335 \\
    66& 4.89770 & 4.89955 & 0.038 & 4.89976 & 0.042 \\
    67& 4.91976 & 4.92224 & 0.050 & 4.92262 & 0.058 \\
    68& 4.94198 & 4.93395 & 0.163 & 4.93404 & 0.161 \\
    69& 4.97544 & 4.95930 & 0.326 & 4.95855 & 0.341 \\
    70& 5.00234 & 4.98595 & 0.329 & 4.98635 & 0.321 \\
    71& 5.05521 & 5.05121 & 0.079 & 5.05153 & 0.073 \\
    72& 5.06222 & 5.07826 & 0.316 & 5.07871 & 0.325 \\
    73& 5.10734 & 5.10605 & 0.025 & 5.10572 & 0.032 \\
    74& 5.14001 & 5.15663 & 0.322 & 5.15649 & 0.320 \\
    75& 5.17375 & 5.18850 & 0.284 & 5.18928 & 0.299 \\
    76& 5.20031 & 5.19720 & 0.060 & 5.19750 & 0.054 \\
    77& 5.22839 & 5.23149 & 0.059 & 5.23154 & 0.060 \\
    78& 5.24876 & 5.27701 & 0.535 & 5.27763 & 0.547 \\
    79& 5.31483 & 5.33671 & 0.410 & 5.33550 & 0.387 \\
    80& 5.35238 & 5.35144 & 0.018 & 5.35154 & 0.016 \\
    81& 5.36938 & 5.36972 & 0.006 & 5.36912 & 0.005 \\
    82& 5.37790 & 5.40947 & 0.584 & 5.40967 & 0.587 \\
    83& 5.41390 & 5.42943 & 0.286 & 5.42989 & 0.295 \\
    84& 5.43467 & 5.44317 & 0.156 & 5.44304 & 0.154 \\
\end{longtable}

\begin{longtable}{|c|ccc|ccc|}
	\caption{\ute SMI fit results}\\
	\hline
	      & \multicolumn{3}{c|}{300 K}     			   & \multicolumn{3}{c|}{4 K}\\ 
		\cline{1-7}
	Index & $f_{exp}$ (MHz) & $f_{calc}$ (MHz) & diff (\%) & $f_{exp}$ (MHz) & $f_{calc}$ (MHz) & diff (\%) \\ 
	\hline
	\endfirsthead
	\multicolumn{7}{r}{Table 6 continued.}\\
	\hline
		  & \multicolumn{3}{c|}{300 K}     			   & \multicolumn{3}{c|}{4 K}\\ 
  			\cline{1-7}
	Index & $f_{exp}$ (MHz) & $f_{calc}$ (MHz) & diff (\%) & $f_{exp}$ (MHz) & $f_{calc}$ (MHz) & diff (\%) \\ 
	\hline
	\endhead
	\hline
	\multicolumn{7}{r}{Table 6 continued on next page.}
	\endfoot
	\hline
	\endlastfoot
	1  & 0.64645 & 0.64254 & -     & 0.66344 & 0.65765 & -    \\
    2  & 0.80333 & 0.80092 & -     & 0.81724 & 0.81425 & -    \\
    3  & 0.86443 & 0.85869 & -     & 0.87702 & 0.87984 & -    \\
    4  & 1.09564 & 1.09407 & 0.143 & 1.11952 & 1.12080 & 0.114\\
    5  & 1.18679 & 1.18992 & 0.263 & 1.21970 & 1.21818 & 0.125\\
    6  & 1.32441 & 1.32471 & 0.023 & 1.34208 & 1.34680 & 0.350\\
    7  & 1.34023 & 1.33357 & 0.499 & 1.36327 & 1.35737 & 0.435\\
    8  & 1.49544 & 1.49019 & 0.353 & 1.51789 & 1.51792 & 0.002\\
    9  & 1.56441 & 1.57149 & 0.450 & 1.59618 & 1.60073 & 0.284\\
    10 & 1.64031 & 1.64096 & 0.040 & 1.67269 & 1.67216 & 0.032\\
    11 & 1.67848 & 1.67996 & 0.088 & 1.71445 & 1.71359 & 0.050\\
    12 & 1.71675 & 1.71182 & 0.288 & 1.75060 & 1.74551 & 0.291\\
    13 & 1.77448 & 1.77473 & 0.014 & 1.80806 & 1.80856 & 0.028\\
    14 & 1.83771 & 1.83767 & 0.002 & 1.87151 & 1.87196 & 0.024\\
    15 & 1.86861 & 1.86685 & 0.094 & 1.90573 & 1.90183 & 0.205\\
    16 & 1.93964 & 1.93352 & 0.317 & 1.97323 & 1.97067 & 0.130\\
    17 & 2.02902 & 2.02240 & 0.327 & 2.06927 & 2.06282 & 0.313\\
    18 & 2.06086 & 2.06388 & 0.146 & 2.09666 & 2.10371 & 0.335\\
    19 & 2.13401 & 2.13646 & 0.115 & 2.17374 & 2.17693 & 0.147\\
    20 & 2.18310 & 2.17778 & 0.244 & 2.23096 & 2.22600 & 0.223\\
    21 & 2.21720 & 2.21601 & 0.054 & 2.26583 & 2.25830 & 0.334\\
    22 & 2.25887 & 2.25321 & 0.251 & 2.30249 & 2.29602 & 0.282\\
    23 & 2.27636 & 2.28024 & 0.170 & 2.32310 & 2.32435 & 0.054\\
    24 & 2.33423 & 2.33367 & 0.024 & 2.37609 & 2.37215 & 0.166\\
    25 & 2.35896 & 2.35370 & 0.223 & 2.40235 & 2.40166 & 0.029\\
    26 & 2.37195 & 2.37131 & 0.027 & 2.41002 & 2.41040 & 0.016\\
    27 & 2.40361 & 2.39962 & 0.166 & 2.46079 & 2.45405 & 0.275\\
    28 & 2.45741 & 2.45189 & 0.225 & 2.50592 & 2.50395 & 0.079\\
    29 & 2.47653 & 2.47529 & 0.050 & 2.51386 & 2.52139 & 0.299\\
    30 & 2.49726 & 2.49516 & 0.085 & 2.55527 & 2.55033 & 0.194\\
    31 & 2.53119 & 2.53407 & 0.113 & 2.58793 & 2.58510 & 0.109\\
    32 & 2.56841 & 2.57293 & 0.176 & 2.61987 & 2.62518 & 0.202\\
    33 & 2.58945 & 2.58640 & 0.118 & 2.64506 & 2.64006 & 0.189\\
    34 & 2.67652 & 2.67401 & 0.094 & 2.72662 & 2.72716 & 0.020\\
    35 & 2.69934 & 2.70106 & 0.064 & 2.75414 & 2.75458 & 0.016\\
    36 & 2.70867 & 2.71641 & 0.285 & 2.76160 & 2.76697 & 0.194\\
    37 & 2.75772 & 2.76163 & 0.142 & 2.81381 & 2.81674 & 0.104\\
    38 & 2.80315 & 2.80522 & 0.073 & 2.85929 & 2.86408 & 0.167\\
    39 & 2.83790 & 2.83401 & 0.137 & 2.88995 & 2.88518 & 0.165\\
    40 & 2.84413 & 2.84397 & 0.006 & 2.89806 & 2.89733 & 0.025\\
    41 & 2.85740 & 2.85678 & 0.022 & 2.91031 & 2.91158 & 0.044\\
    42 & 2.89809 & 2.90021 & 0.073 & 2.95404 & 2.95353 & 0.017\\
    43 & 2.91204 & 2.90687 & 0.178 & 2.96297 & 2.95926 & 0.125\\
    44 & 2.93294 & 2.93629 & 0.114 & 2.98892 & 2.98868 & 0.008\\
    45 & 2.97064 & 2.96887 & 0.060 & 3.03454 & 3.03881 & 0.140\\
    46 & 2.98750 & 2.98376 & 0.125 & 3.04986 & 3.04773 & 0.070\\
    47 & 3.04099 & 3.04359 & 0.086 & 3.10384 & 3.10431 & 0.015\\
    48 & 3.05451 & 3.05513 & 0.020 & 3.11878 & 3.11603 & 0.088\\
    49 & 3.07129 & 3.06617 & 0.167 & 3.12869 & 3.12381 & 0.156\\
    50 & 3.11543 & 3.11631 & 0.028 & 3.17516 & 3.18070 & 0.174\\
    51 & 3.14348 & 3.14681 & 0.106 & 3.21205 & 3.20841 & 0.113\\
    52 & 3.16609 & 3.16333 & 0.087 & 3.23373 & 3.23001 & 0.115\\
    53 & 3.19312 & 3.19392 & 0.025 & 3.25821 & 3.25521 & 0.092\\
    54 & 3.21950 & 3.22107 & 0.049 & 3.28885 & 3.29187 & 0.092\\
    55 & 3.26064 & 3.26213 & 0.046 & 3.32549 & 3.32873 & 0.097\\
    56 & 3.27367 & 3.27710 & 0.105 & 3.33381 & 3.33748 & 0.110\\
    57 & 3.28431 & 3.28793 & 0.110 & 3.35277 & 3.35414 & 0.041\\
    58 & 3.33902 & 3.33560 & 0.103 & 3.39969 & 3.39764 & 0.060\\
    59 & 3.37665 & 3.38210 & 0.161 & 3.44137 & 3.44067 & 0.020\\
    60 & 3.38263 & 3.38841 & 0.171 & 3.44466 & 3.45563 & 0.317\\
    61 & 3.40724 & 3.40264 & 0.135 & 3.47399 & 3.46879 & 0.150\\
    62 & 3.42588 & 3.41984 & 0.176 & 3.48688 & 3.48075 & 0.176\\
    63 & 3.46878 & 3.46450 & 0.123 & 3.52655 & 3.52462 & 0.055\\
    64 & 3.48772 & 3.48306 & 0.134 & 3.55491 & 3.55194 & 0.084\\
    65 & 3.49884 & 3.50575 & 0.197 & 3.56459 & 3.57159 & 0.196\\
    66 & 3.51096 & 3.51080 & 0.005 & 3.58054 & 3.57953 & 0.028\\
    67 & 3.54860 & 3.55098 & 0.067 & 3.61868 & 3.62048 & 0.050\\
    68 & 3.56768 & 3.57417 & 0.182 & 3.64321 & 3.64728 & 0.112\\
    69 & 3.57963 & 3.58351 & 0.108 & 3.65177 & 3.65702 & 0.144\\
    70 & 3.59151 & 3.59154 & 0.001 & 3.66315 & 3.66157 & 0.043\\
    71 & 3.63037 & 3.63146 & 0.030 & 3.70535 & 3.70318 & 0.059\\
    72 & 3.66135 & 3.66185 & 0.014 & 3.73498 & 3.73249 & 0.067\\
    73 & 3.67686 & 3.67367 & 0.087 & 3.75129 & 3.74617 & 0.137\\
    74 & 3.70093 & 3.69729 & 0.098 & 3.77215 & 3.77270 & 0.015\\
    75 & 3.72551 & 3.71979 & 0.154 & 3.79478 & 3.78954 & 0.138\\
    76 & 3.74072 & 3.74038 & 0.009 & 3.81853 & 3.81297 & 0.146\\
    77 & 3.75135 & 3.75430 & 0.079 & 3.82596 & 3.82515 & 0.021\\
    78 & 3.76556 & 3.76595 & 0.010 & 3.83940 & 3.83786 & 0.040\\
    79 & 3.77374 & 3.77640 & 0.071 & 3.84028 & 3.84954 & 0.240\\
    80 & 3.81529 & 3.81717 & 0.049 & 3.89424 & 3.89363 & 0.016\\
    81 & 3.84139 & 3.83861 & 0.072 & 3.91424 & 3.91245 & 0.046\\
    82 & 3.85843 & 3.85711 & 0.034 & 3.93674 & 3.93303 & 0.094\\
    83 & 3.87490 & 3.88271 & 0.201 & 3.94923 & 3.95735 & 0.205\\
    84 & 3.89786 & 3.89296 & 0.126 & 3.97082 & 3.96312 & 0.194\\
    85 & 3.90314 & 3.90305 & 0.002 & 3.98043 & 3.98563 & 0.130\\
    86 & 3.91220 & 3.92250 & 0.263 & 3.98665 & 3.99710 & 0.261\\
    87 & 3.94153 & 3.93803 & 0.089 & 4.01442 & 4.01238 & 0.051\\
    88 & 3.99297 & 3.98031 & 0.318 & 4.06185 & 4.05116 & 0.264\\
    89 & 3.99436 & 3.99491 & 0.014 & 4.07113 & 4.06908 & 0.050\\
    90 & 4.00126 & 4.01160 & 0.258 & 4.07942 & 4.08544 & 0.147\\
    91 & 4.03760 & 4.03141 & 0.154 & 4.11408 & 4.10575 & 0.203\\
    92 & 4.07137 & 4.07176 & 0.010 & 4.15105 & 4.14854 & 0.061\\
    93 & 4.08673 & 4.08572 & 0.025 & 4.16596 & 4.16925 & 0.079\\
    94 & 4.11792 & 4.11397 & 0.096 & 4.19784 & 4.19441 & 0.082\\
    95 & 4.12896 & 4.13582 & 0.166 & 4.20427 & 4.21294 & 0.206\\
    96 & 4.14529 & 4.15212 & 0.164 & 4.22149 & 4.22953 & 0.190\\
    97 & 4.15242 & 4.15367 & 0.030 & 4.23333 & 4.23415 & 0.019\\
    98 & 4.17297 & 4.17074 & 0.053 & 4.24888 & 4.24690 & 0.047\\
    99 & 4.18994 & 4.18888 & 0.025 & 4.27899 & 4.27740 & 0.037\\
    100& 4.20373 & 4.21285 & 0.216 & 4.28904 & 4.29176 & 0.064\\
    101& 4.22958 & 4.23006 & 0.011 & 4.30977 & 4.31112 & 0.031\\
    102& 4.24199 & 4.24296 & 0.023 & 4.32438 & 4.32504 & 0.015\\
    103& 4.25360 & 4.25075 & 0.067 & 4.33687 & 4.33486 & 0.046\\
    104& 4.28642 & 4.28345 & 0.069 & 4.36950 & 4.36567 & 0.088\\
    105& 4.30657 & 4.30624 & 0.008 & 4.39228 & 4.39207 & 0.005\\
    106& 4.31547 & 4.31995 & 0.104 & 4.40638 & 4.40694 & 0.013\\
    107& 4.34118 & 4.34193 & 0.017 & 4.42091 & 4.42370 & 0.063\\
    108& 4.34935 & 4.35598 & 0.152 & 4.43621 & 4.44172 & 0.124\\
    109& 4.36295 & 4.36176 & 0.027 & 4.44922 & 4.44494 & 0.096\\
    110& 4.37974 & 4.37984 & 0.002 & 4.46691 & 4.46202 & 0.110\\
    111& 4.39652 & 4.38614 & 0.237 & 4.47881 & 4.46618 & 0.283\\
    112& 4.40482 & 4.41073 & 0.134 & 4.48982 & 4.49513 & 0.118\\
    113& 4.42003 & 4.41888 & 0.026 & 4.49809 & 4.50730 & 0.204\\
    114& 4.42672 & 4.42968 & 0.067 & 4.51862 & 4.51611 & 0.055\\
    115& 4.44940 & 4.46193 & 0.281 & 4.53519 & 4.54067 & 0.121\\
    116& 4.46937 & 4.47430 & 0.110 & 4.55608 & 4.55307 & 0.066\\
    117& 4.48235 & 4.48380 & 0.032 & 4.56374 & 4.56785 & 0.090\\
    118& 4.50482 & 4.50569 & 0.019 & 4.59148 & 4.59609 & 0.100\\
    119& 4.50733 & 4.51327 & 0.132 & 4.59729 & 4.60078 & 0.076\\
    120& 4.52413 & 4.52782 & 0.081 & 4.60960 & 4.61562 & 0.131\\
    121& 4.53911 & 4.54276 & 0.080 & 4.62762 & 4.63458 & 0.150\\
    122& 4.56333 & 4.56602 & 0.059 & 4.64820 & 4.65923 & 0.237\\
    123& 4.57367 & 4.57286 & 0.018 & 4.66121 & 4.66642 & 0.112\\
    124& 4.60540 & 4.60303 & 0.051 & 4.69278 & 4.69228 & 0.011\\
    125& 4.61532 & 4.62032 & 0.108 & 4.70740 & 4.71043 & 0.064\\
    126& 4.62818 & 4.62519 & 0.065 & 4.71517 & 4.71609 & 0.020\\
    127& 4.63514 & 4.63026 & 0.105 & 4.72725 & 4.71991 & 0.155\\
    128& 4.64194 & 4.64199 & 0.001 & 4.73331 & 4.73754 & 0.089\\
    129& 4.66690 & 4.66430 & 0.056 & 4.75924 & 4.75618 & 0.064\\
    130& 4.66790 & 4.66597 & 0.041 & 4.76674 & 4.75845 & 0.174\\
    131& 4.69246 & 4.68730 & 0.110 & 4.78743 & 4.78038 & 0.147\\
    132& 4.70527 & 4.71080 & 0.118 & 4.80128 & 4.80542 & 0.086\\
    133& 4.72046 & 4.73008 & 0.203 & 4.81305 & 4.82061 & 0.157\\
    134& 4.75051 & 4.75395 & 0.072 & 4.84664 & 4.84559 & 0.022\\
    135& 4.75900 & 4.76169 & 0.057 & 4.85021 & 4.85330 & 0.064\\
    136& 4.76976 & 4.77384 & 0.085 & 4.86402 & 4.86321 & 0.017\\
    137& 4.77918 & 4.77950 & 0.007 & 4.87531 & 4.87473 & 0.012\\
    138& 4.79086 & 4.79493 & 0.085 & 4.88149 & 4.88582 & 0.089\\
    139& 4.80450 & 4.80244 & 0.043 & 4.89785 & 4.89637 & 0.030\\
    140& 4.81875 & 4.81484 & 0.081 & 4.91309 & 4.90863 & 0.091\\
    141&         &         &       & 4.93822 & 4.94346 & 0.106\\
    142&         &         &       & 4.95627 & 4.95918 & 0.059\\
    143&         &         &       & 4.96168 & 4.96342 & 0.035\\
    144&         &         &       & 4.98300 & 4.98143 & 0.032\\
    145&         &         &       & 4.98846 & 4.99420 & 0.115\\
    146&         &         &       & 4.99438 & 5.00516 & 0.215\\
    147&         &         &       & 5.01215 & 5.01596 & 0.076\\
    148&         &         &       & 5.02882 & 5.03051 & 0.034\\
    149&         &         &       & 5.04591 & 5.05275 & 0.135\\
    150&         &         &       & 5.06142 & 5.06270 & 0.025\\
\end{longtable}	

% \bibliography{literature}

% \end{document}

\end{document}